\documentclass[final,nopreprintline, 3p,times,9pt]{elsarticle}
\biboptions{numbers,sort&compress}
\journal{}
\usepackage[T1]{fontenc} 
\usepackage{amssymb}
\usepackage{amsmath}
\usepackage{color}
\usepackage{xspace}
\usepackage{appendix}
\usepackage{float}
\usepackage{hyperref}
\usepackage{multirow}
\usepackage{diagbox}
\usepackage[utf8x]{inputenc}
\usepackage{soul}
\usepackage[compat=1.1.0]{tikz-feynhand}
\usepackage{subfig}
\usepackage{kantlipsum}

\newcommand{\nn}{\nonumber}
\newcommand{\bea}{\begin{align}}
\newcommand{\eea}{\end{align}}
\newcommand{\beq}{\begin{equation}}
\newcommand{\eeq}{\end{equation}}
\newcommand{\bqa}{\begin{eqnarray}}
\newcommand{\eqa}{\end{eqnarray}}

\newcommand{\ep}{\epsilon}

\allowdisplaybreaks[4]

\begin{document}

\begin{frontmatter}

\title{Complete two-loop QCD amplitudes for $tW$ production at hadron colliders}
\author[1]{Long-Bin Chen}
\author[2]{Liang Dong}
\author[2]{Hai Tao Li}
\author[3,4,5]{Zhao Li}
\author[2]{Jian Wang}
\author[2]{Yefan Wang}

\address[1]{School of Physics and Materials Science, Guangzhou University, Guangzhou 510006, China}
\address[2]{School of Physics, Shandong University, Jinan, Shandong 250100, China}
\address[3]{Institute of High Energy Physics, Chinese Academy of Sciences, Beijing 100049, China}
\address[4]{School of Physics Sciences, University of Chinese Academy of Sciences, Beijing 100039, China}
\address[5]{Center for High Energy Physics, Peking University, Beijing 100871, China}

\begin{abstract}
We calculate the complete two-loop QCD amplitudes for hadronic $tW$ production by combining analytical and numerical techniques. 
The  amplitudes have been first reduced to master integrals of eight planar and seven non-planar families, which can contain at most four massive propagators.
Then a rational transformation of the master integrals is found to obtain  a good basis so that the dimensional parameter  decouples from the kinematic variables in the denominators of reduction coefficients. 
The master integrals are computed by solving their differential equations numerically.
We find that the finite part of the two-loop squared amplitude is stable in the bulk of the phase space.
After phase space integration and convolution with the parton distributions, it increases the LO cross section by about $3\%$.

\end{abstract}


\end{frontmatter}

\section{Introduction}
\label{sec:intro}

The top quark, first discovered at the Fermilab Tevatron \cite{D0:1995jca,CDF:1995wbb}, is the heaviest elementary particle in the Standard Model (SM),
and  may have a close relation to the electroweak symmetry breaking due to its large coupling with the Higgs boson.
The single top-quark production processes deserve detailed studies because they can be used to measure the $Wtb$ coupling structure, which may be modified by new physics.
Moreover,  they are often considered as important backgrounds to many new physics searches.
In this paper, we focus on the $tW$ associated production, which has the second largest rate, after the t-channel, at the large hadron collider (LHC).
The inclusive and differential cross sections have been measured to an accuracy of $10\%$ \cite{ATLAS:2016ofl,ATLAS:2017quy,CMS:2021vqm,CMS:2022ytw}.

In order to provide precise theoretical predictions for the rate and kinematical distributions, higher-order quantum corrections are indispensable.
The next-to-leading order (NLO) quantum chromodynamics (QCD) correction 
has been obtained  for stable $tW$ production \cite{Giele:1995kr,Zhu:2001hw,Cao:2008af, Kant:2014oha}.
The correction to the process including decays was calculated in \cite{Campbell:2005bb}.
The NLO QCD result was interfaced with  parton shower within both the MC@NLO and POWHEG  formalisms \cite{Frixione:2008yi,Re:2010bp,Jezo:2016ujg}.
The NLO electroweak correction was computed in \cite{Beccaria:2007tc}.

There have been also some efforts devoted to the calculation of the corrections beyond the NLO in QCD.
The approximate next-to-next-to-next-to-leading order total cross section was obtained by expanding the threshold resummation formula 
\cite{Kidonakis:2006bu,Kidonakis:2010ux,Kidonakis:2016sjf,Kidonakis:2021vob}.
The  corrections induced by soft gluons have been resummed to all orders in the strong coupling  \cite{Li:2019dhg}.
The N-jettiness soft function, which is one of the ingredients in a full NNLO QCD correction, has been calculated by two of the authors \cite{Li:2016tvb,Li:2018tsq}.
Recently, we reported the analytical results of the leading color and light quark-loop contributions to the NNLO virtual corrections \cite{Chen:2022yni} using the two-loop master integrals obtained in \cite{Chen:2021gjv,Long:2021vse,Wang:2022enl}.
We have investigated the dominance of the leading color result in the one-loop squared amplitudes, of which the full color result was computed analytically in \cite{Chen:2022ntw}.
However, it is ideal to have the full two-loop virtual correction, which is the aim of this paper.
This is made possible due to the impressive progresses in both the analytical  and numerical  methods of the computation of Feynman diagrams; see, e.g.,  \cite{Henn:2013pwa,Liu:2017jxz}.

This paper is organised as follows. In section~\ref{sec:bare}, we describe the basic setup and details in our  calculation of complete two-loop bare amplitudes. 
We also discuss the methods to cancel the ultra-violet (UV) and infra-red (IR) divergences in the bare amplitudes. 
The finite part of the squared amplitude defines
the hard function which is needed in a NNLO calculation.  
The numerical results for the NNLO hard function are presented in section~\ref{sec:num}. 
Finally we make the conclusion in section~\ref{sec:concl}.

\section{Calculation method}
\label{sec:bare}

We calculate the two-loop corrections to the process $g(k_1)+b(k_2)\rightarrow W(k_3)+t(k_4)$ with  kinematical conditions  $k_1^2=k_2^2=0$, $k_3^2 = m_W^2$ and $k_4^2 = (k_1+k_2-k_3)^2=m_t^2$.
The Lorentz invariant Mandelstam variables are defined by
\begin{align}
	s=(k_1+k_2)^2\,, \qquad t=(k_1-k_3)^2\,, \qquad u=(k_2-k_3)^2 \,,
	\label{kine}
\end{align}
so that $s+t+u=m_W^2+m_t^2$.

The amplitude is expanded in a series of $\alpha_s$,
\begin{align}
     \mathcal{M} = \sum_{i=0}^\infty\left(\frac{\alpha_s}{4\pi}\right)^i \mathcal{M}^{(i)} \,,
\end{align}
where $\alpha_s$ is the strong coupling.
In this work, we do not keep the polarization information and 
focus only on amplitude squared.
We have presented the analytic result of the one-loop squared amplitude in \cite{Chen:2022ntw}.
In this paper, we calculate the interference between the two-loop and tree-level amplitudes.
According to the color structure and fermion-loop contribution, it is decomposed as 
\bqa 
\label{eq:amp}
\mathcal{A}^{(2)}&=&2\; {\rm Re }\sum_{\text{spin,color}}\mathcal{M}^{(0)*} \mathcal{M}^{(2)}\nonumber\\
&=&N_C^4 A+ N_C^2 B+ C+\frac{1}{N_C^2}D + n_l\left(N_C^3 E_l+N_C F_l +\frac{1}{N_C}G_l \right)
+n_h\left(N_C^3 E_h+N_C F_h +\frac{1}{N_C}G_h \right) \,,
\eqa
where $N_C$ is the number of  quark colors, $n_l$ ($n_h$) is the total number of massless (massive) quark flavors.
The coefficients of various color structures are denoted by 
 $A,B,C,D,E,F,G$. 
In our previous paper \cite{Chen:2022yni}, we have obtained the analytical results of the leading color $A$ and light fermion-loop contributions $E_l,F_l$ and $G_l$. 
In this work, we continue  to calculate  the remaining contributions $B,C,D, E_h,F_h$ and $ G_h$.

\begin{figure}[H]
	\centering
	\begin{minipage}{0.2\linewidth}
		\centering
		\includegraphics[width=1\linewidth]{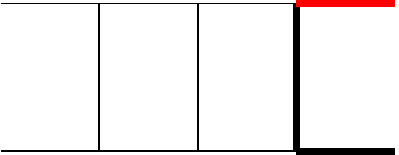}
		\caption*{P1}
	\end{minipage}
	\begin{minipage}{0.2\linewidth}
		\centering
		\includegraphics[width=1\linewidth]{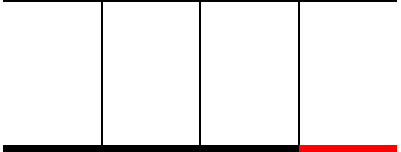}
		\caption*{P2}
	\end{minipage}
	\begin{minipage}{0.2\linewidth}
		\centering
		\includegraphics[width=1\linewidth]{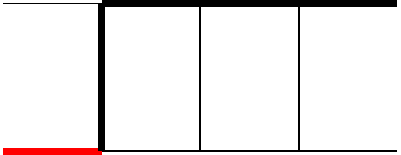}
		\caption*{P3}
	\end{minipage}
	\begin{minipage}{0.2\linewidth}
		\centering
		\includegraphics[width=1\linewidth]{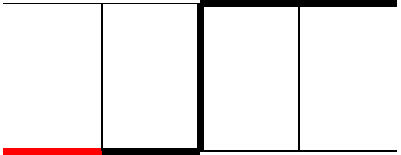}
		\caption*{P4}
	\end{minipage}
	\begin{minipage}{0.2\linewidth}
		\centering
		\includegraphics[width=1\linewidth]{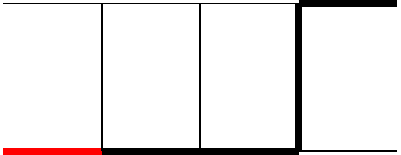}
		\caption*{P5}
	\end{minipage}
	\begin{minipage}{0.2\linewidth}
		\centering
		\includegraphics[width=1\linewidth]{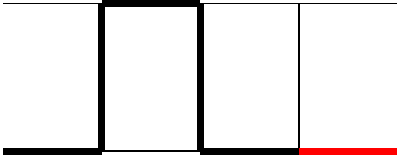}
		\caption*{P6}
	\end{minipage}
	\begin{minipage}{0.2\linewidth}
	\centering
	\includegraphics[width=1\linewidth]{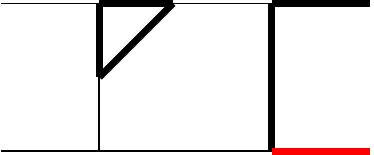}
	\caption*{P7}
\end{minipage}
	\begin{minipage}{0.2\linewidth}
	\centering
	\includegraphics[width=1\linewidth]{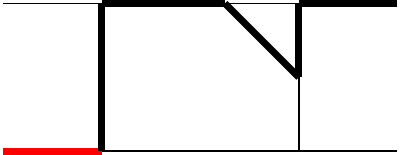}
	\caption*{P8}
\end{minipage}
	\caption{Planar integral topologies for $gb\to tW$. The black and red thick lines represent the top quark and the $W$ boson, respectively. The black thin lines denote massless particles. For the P1 and P2 topologies, the  symmetric  diagrams $(k_1 \leftrightarrow k_2)$ are needed in the amplitude. }
	\label{fig:P}
\end{figure}

\begin{figure}[H]
	\centering
\begin{minipage}{0.20\linewidth}
		\centering
		\includegraphics[width=1\linewidth]{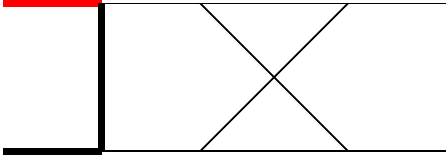}
		\caption*{NP1}
\end{minipage}
\begin{minipage}{0.20\linewidth}
	\centering
	\includegraphics[width=1\linewidth]{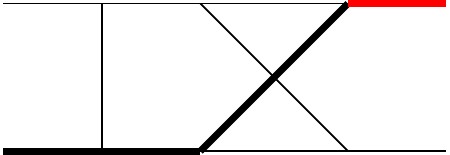}
	\caption*{NP2}
\end{minipage}
\begin{minipage}{0.20\linewidth}
	\centering
	\includegraphics[width=1\linewidth]{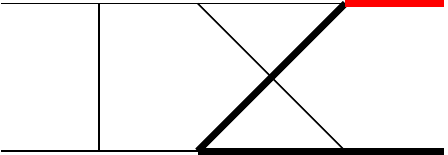}
	\caption*{NP3}
\end{minipage}
\begin{minipage}{0.20\linewidth}
	\centering
	\includegraphics[width=1\linewidth]{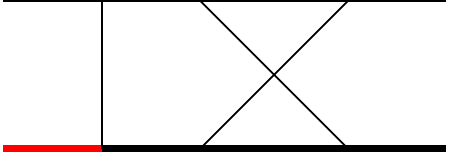}
	\caption*{NP4}
\end{minipage}
\begin{minipage}{0.20\linewidth}
	\centering
	\includegraphics[width=1\linewidth]{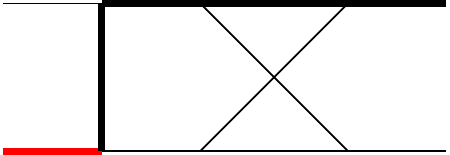}
	\caption*{NP5}
\end{minipage}
\begin{minipage}{0.20\linewidth}
	\centering
	\includegraphics[width=1\linewidth]{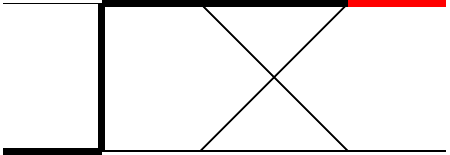}
	\caption*{NP6}
\end{minipage}
\begin{minipage}{0.20\linewidth}
	\centering
	\includegraphics[width=1\linewidth]{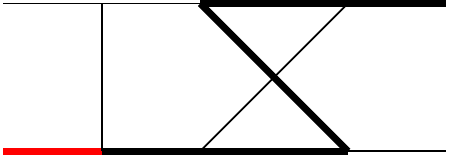}
	\caption*{NP7}
\end{minipage}
\caption{Non-planar integral topologies for $gb\to tW$. The black and red thick lines represent the top quark and the $W$ boson, respectively. The black thin lines denote massless particles. For the NP2 and NP4 topologies, the  symmetric  diagrams $(k_1 \leftrightarrow k_2)$ are needed in the amplitude. }
\label{fig:NP}
\end{figure}

We generate the two-loop Feynman diagrams  using {\tt FeynArts} \cite{Hahn:2000kx} and perform the Dirac algebra with the help of {\tt FeynCalc} \cite{Shtabovenko:2020gxv}.
In our calculation, the anticommuting $\gamma_5$ scheme proposed in \cite{Korner:1991sx} is implemented. 
The traces with two $\gamma_5$'s are easy, while
the traces with one $\gamma_5$ vanish because there are only three independent momenta in the $tW$ production process.
More discussion on the implementation  can be found in \cite{Chen:2022ntw}.
After the spin and polarization summation, all Lorentz indices are contracted, 
and the squared amplitude is written as a combination of  a large number of scalar integrals. They are reduced to a set of basis integrals, called master integrals,  making use of the integration-by-part (IBP) identities,
which are automatically generated in {\tt FIRE} \cite{Smirnov:2019qkx}. 
It turns out that all the scalar integrals could be expressed by 15 families of master integrals \footnote{Some master integrals that can be factorized as two one-loop integrals are  easy to calculate and not considered here.},
which are defined according to the topologies of the Feynman integrals.
Specifically, the master integrals are categorized in terms of eight planar and seven non-planar topologies, which are shown in figures \ref{fig:P} and \ref{fig:NP}, respectively. 
They have up to four massive propagators, which makes the analytical structure complicated.


\begin{table*}[ht]
    \centering
    \begin{tabular}{|c|c|p{0.7 \linewidth}|}
        \hline \hline
        \multicolumn{2}{|c|}{\textbf{Name}} & \multicolumn{1}{|c|}{\textbf{Definition}} \\
        \hline \hline \multirow{8}{*}{Planar}

       &  P1 &
        $
            q_{1}^{2},
            q_{2}^{2},
            (q_{1} - k_{1})^{2},
            (q_{1} + k_{2})^{2},
            (q_{1} + q_{2} - k_{1})^{2}, 
                   $
        \newline
        $
            (q_{2} - k_{1} - k_{2})^{2},
            (q_{2}-k_{3})^{2} - m_{t}^{2},
            (q_{1}+k_{4})^{2} - m_{t}^{2},
            (q_{2} - k_{1})^{2}
        $
        \\
        \cline{2-3}
       &  P2 &
        $
            q_{1}^{2},
            q_{2}^{2},
            (q_{1} - k_{2})^{2},
            (q_{1} - k_{3})^{2}-m_t^2,
            (q_{1}+q_{2} - k_{2})^{2},
        $
        \newline
        $
            (q_{2}  + k_{1})^{2} ,
            (q_{2}  - k_{2}+ k_{3})^{2} - m_{t}^{2},
            (q_{1}  - k_{1}- k_{2})^{2},
            (q_{2}  + k_{1}- k_{2})^{2}
        $
        \\
          \cline{2-3}
       &  P3 &
        $
            q_{1}^{2}- m_{t}^{2},
            q_{2}^{2}- m_{t}^{2},
            (q_{1} - k_{1})^{2}- m_{t}^{2},
            (q_{1} - k_{3})^{2},
            (q_{1}+q_{2} - k_{1})^{2},
        $
        \newline
        $
            (q_{2} - k_{4})^{2},
            (q_{2} - k_{1} + k_{3})^{2},
            (q_{2} - k_{2} + k_{3})^{2},
            (q_{1} + k_{2} - k_{3})^{2}
        $
        \\
          \cline{2-3}
       &  P4 &
        $
            q_{1}^{2},
            q_{2}^{2}- m_{t}^{2},
            (q_{1} - k_{1})^{2},
            (q_{1} - k_{4})^{2}- m_{t}^{2},
            (q_{1} + q_{2} - k_{1})^{2} - m_{t}^{2},
        $
        \newline
        $
            (q_{2} + k_{2} - k_{3} )^{2},
            (q_{2} - k_{3} )^{2},
            (q_{1} + k_{3}- k_{1} )^{2},
            (q_{2} - k_{3}+ k_{1} )^{2} 
        $
        \\
          \cline{2-3}
       &  P5 &
        $
            q_{1}^{2}- m_{t}^{2},
            q_{2}^{2}- m_{t}^{2},
            (q_{1} - k_{1})^{2}- m_{t}^{2},
            (q_{1} - k_{4})^{2},
            (q_{1} + q_{2} - k_{1})^{2},
        $
        \newline
        $
            (q_{2} + k_{2} - k_{3})^{2},
            (q_{2} - k_{3})^{2},
            (q_{2} + k_{4})^{2},
            (q_{1} - k_{1} + k_{3})^{2}
        $
        \\
          \cline{2-3}
        & P6 &
        $
            q_{1}^{2}- m_{t}^{2},
            q_{2}^{2},
            (q_{1} - k_{1})^{2}- m_{t}^{2},
            (q_{1} - k_{4})^{2},
            (q_{1} + q_{2} - k_{1})^{2}- m_{t}^{2},
        $
        \newline
        $
            (q_{2} + k_{2})^{2},
            (q_{2} + k_{2} - k_{3})^{2}-m_t^2,
            (q_{2} - k_{1})^{2},
            (q_{2} - k_{1} + k_{3})^{2}
        $
        \\
        \cline{2-3}
        & P7 &
        $
             q_{1}^{2} - m_{t}^{2},
             q_{2}^{2},
            (q_{1} + q_{2})^{2} - m_{t}^{2},
            (q_{1} + k_{1} )^{2} - m_{t}^{2},
            (q_{2} - k_{1})^{2} ,
        $
        \newline
        $
            (q_{2} - k_{1} - k_{2} )^{2} ,
            (q_{2} - k_{4})^{2} - m_{t}^{2},
            (q_{1} - k_{2})^{2},
            (q_{1} - k_{3})^{2}
        $
         \\
        \cline{2-3}
       &  P8 &
        $
             q_{1}^{2} - m_{t}^{2},
             q_{2}^{2},
            (q_{1} - k_{1} )^{2} - m_{t}^{2},
            (q_{1} + k_{2} - k_{3})^{2},
            (q_{1} - k_{3})^{2} ,
        $
        \newline
        $
            (q_{1} + q_{2} - k_{1} )^{2}- m_{t}^{2} ,
            (q_{2} - k_{4})^{2} - m_{t}^{2},
            (q_{1} + q_{2})^{2},
            (q_{2} - k_{2})^{2}
        $
        \\
        \hline \hline \multirow{8}{*}{Non-Planar}

       & NP1 &
        $
            q_{1}^{2},
            (q_{2} - q_{1})^{2},
            q_{2}^{2},
            (q_{1} + k_{1} )^{2},
            (q_{2} - q_{1} + k_{2} )^{2},
        $
        \newline
        $
            (q_{2} + k_{1} + k_{2} )^{2},
            (q_{2} + k_{4} )^{2} - m_{t}^{2},
            (q_{1} - k_{3} )^{2},
            (q_{2} + k_{1} )^{2}
        $
        \\
        \cline{2-3}
       &  NP2 &
        $
            q_{1}^{2},
            q_{2}^{2} - m_{t}^{2},
            (q_{1} + q_{2})^{2} - m_{t}^{2},
            (q_{1} + k_{1})^{2},
            (q_{1} + q_{2} + k_{4})^{2},
        $
        \newline
        $
            (q_{1} + q_{2} + k_{1} - k_{3})^{2},
            (q_{2} - k_{3})^{2},
            (q_{1} - k_{2})^{2},
             (q_{1} + q_{2} + k_{2} - k_{3})^{2}
        $
        \\
        \cline{2-3}
        & NP3 &
        $
            q_{1}^{2},
            q_{2}^{2},
            (q_{1} - k_{1})^{2},
            (q_{1} + k_{2})^{2},
            (q_{1} + q_{2} - k_{1})^{2},
        $
        \newline
        $
            (q_{1} + q_{2} +k_{2} - k_{3})^{2} - m_{t}^{2},
            (q_{2} - k_{3})^{2} - m_{t}^{2},
            (q_{1} - k_{4})^{2} - m_{t}^{2},
            (q_{2} - k_{1})^{2}
        $
        \\
        \cline{2-3}
        & NP4 &
        $
            q_{1}^{2},
            q_{2}^{2},
            (q_{1} - k_{1})^{2},
            (q_{1} - k_{3})^{2} - m_{t}^{2},
            (q_{1} + q_{2} - k_{1})^{2},
        $
        \newline
        $
            (q_{2} + k_{2})^{2},
            (q_{1} + q_{2} + k_{2} - k_{3})^{2} - m_{t}^{2},
            (q_{2} - k_{1})^{2},
            (q_{1} - k_{2})^{2}
        $
        \\
        \cline{2-3}
       &  NP5 &
        $
            q_{1}^{2} - m_{t}^{2},
            q_{2}^{2},
            (q_{1} - k_{1})^{2} - m_{t}^{2},
            (q_{1} - k_{3})^{2},
            (q_{1} + q_{2} - k_{1})^{2} - m_{t}^{2},
        $
        \newline
        $
            (q_{2} + k_{2})^{2},
            (q_{1} + q_{2}+ k_{2} - k_{3})^{2} ,
            (q_{1} + q_{2}+ k_{4})^{2} ,
            (q_{1} + k_{2} - k_{3})^{2}
        $
        \\
        \cline{2-3}
       &  NP6 &
        $
            q_{1}^{2} - m_{t}^{2},
            q_{2}^{2},
            (q_{1} - k_{1})^{2} - m_{t}^{2},
            (q_{1} - k_{4})^{2},
            (q_{1} + q_{2} - k_{1})^{2} - m_{t}^{2},
        $
        \newline
        $
            (q_{2} + k_{2})^{2},
            (q_{1} + q_{2}- k_{1} + k_{3})^{2} ,
            (q_{2} - k_{1})^{2} ,
            (q_{1} - k_{1} + k_{3})^{2}
        $
        \\
        \cline{2-3}
      &   NP7 &
        $
            q_{1}^{2}- m_{t}^{2},
            q_{2}^{2}- m_{t}^{2},
            (q_{1} + q_{2})^{2},
            (q_{1} + k_{1})^{2} - m_{t}^{2},
            (q_{1} + q_{2} + k_{4})^{2} - m_{t}^{2},
        $
        \newline
        $
            (q_{2} + k_{2} - k_{3})^{2},
            (q_{2} - k_{3})^{2},
            (q_{1} + k_{2}- k_{4})^{2},
            (q_{2} - k_{4})^{2}
        $
        \\
        \hline \hline
    \end{tabular}
    \caption{Definitions of the master integral families in terms of the denominators $D_i$.
      The above $q_{1}$ and $q_{2}$ are loop momenta while $k_{1}$, $k_{2}$, $k_{3}$  and $k_{4}$
      are external momenta defined in Eq.~\eqref{kine}.
    }
    \label{tab:families}
\end{table*}

The master integrals corresponding to the above topologies are defined as follows:
\bqa
I(n_1,n_2,n_3,n_4,n_5,n_6,n_7,n_8,n_9)=
e^{2\epsilon \gamma_E}\int \frac{\text{d}^d q_1}{i \pi^{d/2}}\frac{\text{d}^d q_2}{i \pi^{d/2}}\frac{D_8^{-n_8} D_9^{-n_9}}{D_1^{n_1} D_2^{n_2} \cdots D_7^{n_7}},
\eqa
where  $d=4-2\ep$ is the spacetime dimension, $\gamma_E\approx 0.5772$ is the Euler-Mascheroni constant, and the denominators $D_i$ in each topology are  listed in table \ref{tab:families}.

The master integrals in the P1, NP1 and P2 topologies have been calculated analytically in \cite{Chen:2021gjv,Long:2021vse}
using the method of differential equations \cite{Kotikov:1990kg,Kotikov:1991pm}. 
There are 31,34 and 38 master integrals in the  P1, NP1 and P2 topologies, respectively. 
And the differential equations have been transformed to the canonical form \cite{Henn:2013pwa}
after constructing a proper basis. 
The solutions can be expressed as multiple polylogarithms \cite{Goncharov:1998kja}. 
The master integrals in the NP2, NP3 and NP4 topologies have also been computed after expansion in $m_W^2$ \cite{Wang:2022enl}.

In the other integral families, however, multiple square roots of three variables appear in the differential equations.
It is challenging, if not impossible, to find a transformation to rationalize all the square roots simultaneously.
Another obstruction is the fact that elliptic integrals are needed in many sectors 
of the integral families.
Despite impressive progress in the past decade \cite{Primo:2016ebd,Primo:2017ipr,Adams:2017tga,Harley:2017qut,Adams:2018bsn,Broedel:2019hyg,Broedel:2019kmn,Frellesvig:2019byn,Walden:2020odh,Muller:2022gec,Pogel:2022yat,Pogel:2022ken,Dlapa:2022wdu}, the approach to analytically compute the Feynman integrals depending on several elliptic curves   is not as full-fledged as that for Feynman integrals which can be evaluated to multiple polylogarithms.
Therefore, we choose to perform the calculation of these integrals numerically  using a method 
that is efficient enough for phenomenology analysis. 

We first identify the master integrals in each integral family, to which the other scalar integrals can be reduced.
It is not required in this process that a canonical basis is chosen, though we have succeeded in obtaining such a basis in some sectors.
We carry out only rational transformation of the basis and demand that the reduction coefficients have a ``good'' factorization property, i.e., the denominators of the coefficients can be written as $N(d + M)$ with $N$  independent of $d$ and $M$ a rational number \cite{Smirnov:2020quc,Usovitsch:2020jrk}.
This is always possible because the integrals should not have poles or branch cuts depending on the spacetime dimension $d$.
The choice of such a kind of ``good'' basis integrals  turns out to be very important to minish the size of the coefficients after IBP  reduction and to solve the differential equations efficiently.
In some cases, choosing an appropriate basis  avoids the cancellation of large numbers and thus makes the numerical calculation more efficient.
We show the master integrals for the most complex  integral families NP5, NP6 and NP7 in the appendix. 

\begin{figure}[ht]
    \centering
    \includegraphics[width=0.35\textwidth]{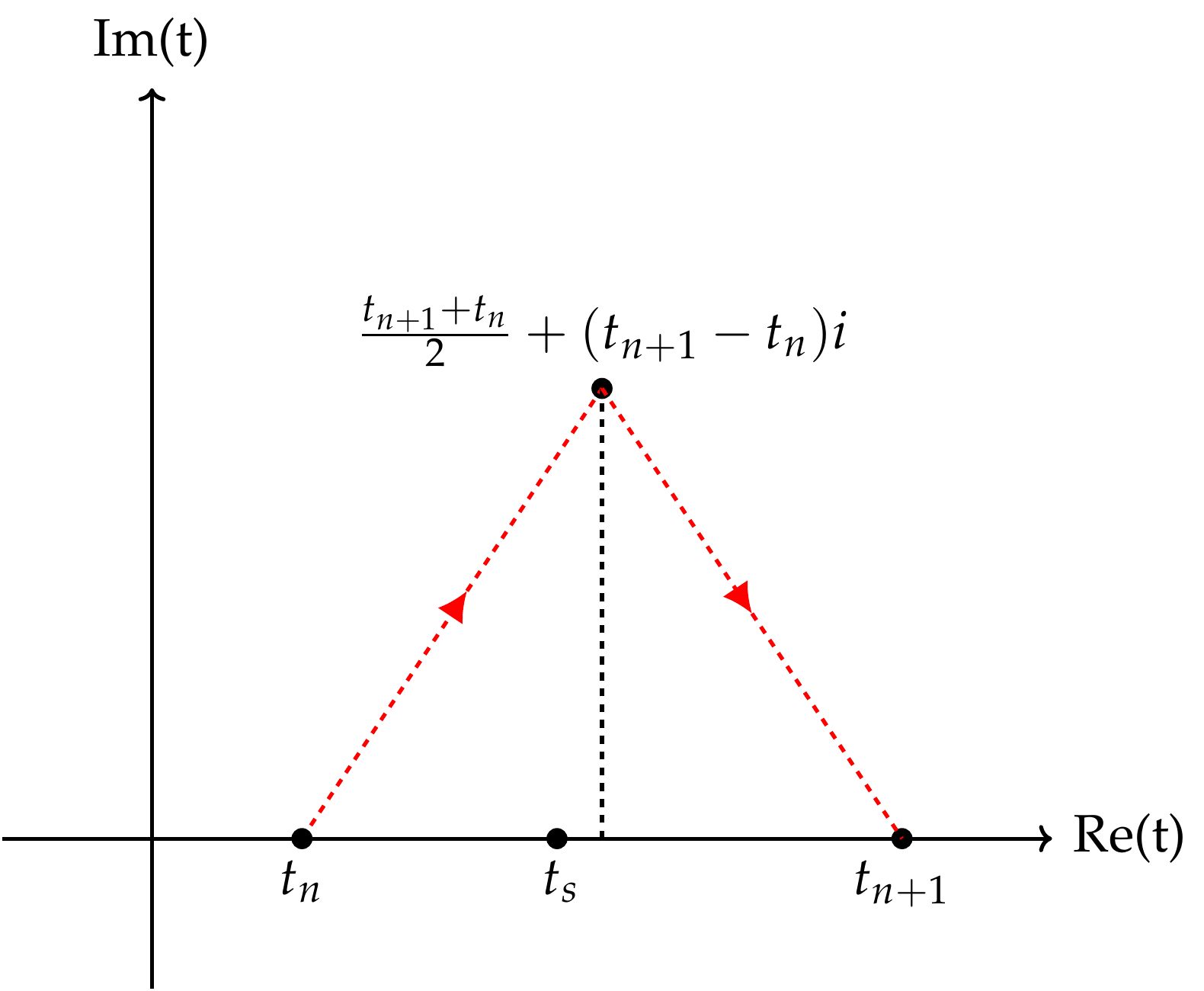}
    \caption{A contour that was set in the numerical evaluation of the master integrals when using the {\tt AMFlow} package. }
    \label{fig:contour}
\end{figure}

These master integrals are computed numerically. 
Specifically, we construct the differential equations of the master integrals with respect to $s$ and $t$. And then, we calculate them at one kinematic point $(s_0,t_0)$, which is chosen in physical region, using the {\tt AMFlow} \cite{Liu:2017jxz,Liu:2022chg} package. 
The results at the other phase space points can be obtained by solving the differential equations in terms of a combination of multiple series expansions. 
We have made use of the {\tt DESolver} function implemented in the {\tt AMFlow} package.
The input includes the set of differential equations as well as an integration path.
There may be pseudo poles on the path, which appear as poles in the differential equations but do not exist in the master integrals.
In order to avoid numerical instability, we choose a contour around the pseudo poles.
A typical example is shown in figure   \ref{fig:contour}.
The integration is performed from $t_n$ to $t_{n+1}$  via a point $\frac{(t_n+t_{n+1})}{2}+(t_{n+1}-t_n)i$ in the complex plane of $t$, given that $t_s\in (t_n, t_{n+1} )$ is a pseudo pole on the real axis.
Notice that it does not matter whether the contour goes above or below the pseudo pole.

We have checked our calculation of the master integrals in different methods.
Firstly, we compare the results of master integrals at a phase space point $(s_i,t_i)$ obtained in two ways. One is from the direct evaluation of the {\tt AMFlow} package, while the other is derived from solving the differential equation as described above.
The numerical agreement between these two methods could  reach more than 50 digits for the integrals of transcendental weight four\footnote{In some cases, we checked this agreement for the integrals up to transcendental weight six.}.
Secondly, the master integrals  have also been computed at some  phase space points using the {\tt FIESTA} \cite{Smirnov:2015mct} package that is developed with the sector decomposition method.
We have found an agreement within the numerical uncertainties of  {\tt FIESTA}.
Thirdly, the numerical results are compared against the analytic ones that are available in some families, such as  the P1, P2 and NP1 topologies \cite{Chen:2021gjv}.

The two-loop amplitudes mentioned above contain both UV and IR divergences. 
The UV divergences cancel against the contribution from renormalization of the couplings, masses and field strength. 
As a result, the renormalized amplitudes can be written as
\begin{align}\label{eq:mren}
    \mathcal{M}_{\rm ren} = Z_{g}^{1/2}Z_{b}^{1/2}Z_{t}^{1/2}\left(\mathcal{M}_{\rm bare}\big|_{\alpha_s^{\rm bare} \to Z_{\alpha_s} \alpha_s;\  m_{t,\rm bare} \to Z_{m} m_t }\right)\;,
\end{align}
where $Z_{g,b,t}$ represent the  renormalization constants for the external states, and $Z_{\alpha_s}$ and $Z_{m}$ are renormalization factors for the strong coupling $\alpha_s$ and the top-quark mass, respectively.  
Their explicit expressions can be found in \cite{Broadhurst:1991fy,Melnikov:2000zc,Czakon:2007wk,Czakon:2007ej}.

The left IR divergences in $ \mathcal{M}_{\rm ren} $ should be combined with those from real corrections to give finite predictions for the cross sections.
Though we have not computed the latter explicitly in this work, the IR divergences can be reconstructed from some universal anomalous dimensions
due to the  understanding of the factorization structure of the amplitudes in the IR limit.
We obtain a finite remainder after subtracting the IR divergences
\begin{align}
    \mathcal{M}_{\rm fin} =  \mathbf{Z}^{-1}\mathcal{M}_{\rm ren},
\label{eq:Mfin}    
\end{align}
where the factor $\mathbf{Z}$ is known  up to two-loop level for general processes in QCD ~\cite{Becher:2009cu,Becher:2009qa,Becher:2009kw,Ferroglia:2009ep,Mitov:2010xw}
and to three-loop level for single top production~\cite{Kidonakis:2019nqa}. 
The explicit expression of this factor for $tW$ production  can be found in our previous papers \cite{Chen:2022ntw,Chen:2022yni}.

We have made nontrivial validations at the amplitude level.
First, we reproduce the leading color result which has been obtained in analytical form in \cite{Chen:2022yni}.
Second, all the IR divergences of the full color result  are indeed canceled out in Eq.(\ref{eq:Mfin}).

We define the squared $ \mathcal{M}_{\rm fin} $ as the hard function that would be used in a NNLO computation,
\begin{align}
    H =\big|  \mathcal{M}_{\rm fin}  \big|^2= H^{(0)} + \frac{\alpha_s}{4\pi} H^{(1)}  +\left(\frac{\alpha_s}{4\pi} \right)^2 H^{(2)}
    +\mathcal{O}(\alpha_s^3)\;,
\end{align}
where we have made a perturbative expansion in the second equation.
Similar to Eq.~(\ref{eq:amp}), the second order of the hard function can be written as
\begin{align}\label{eq:H2}
    H^{(2)} =& N_C^4 H_A + N_C^2 H_B + H_C + \frac{1}{N_C^2} H_D  + n_l \left(N_C^3 H_{El}+N_C H_{Fl} + \frac{1}{N_C} H_{Gl} \right)
         \nn \\   + &   n_h \left(N_C^3 H_{Eh}+N_C H_{Fh} + \frac{1}{N_C} H_{Gh} \right)\;.
\end{align}
We have presented the results of $H_A$ and the terms proportional to $n_l$ in our previous paper \cite{Chen:2022yni}.
In this work, we provide the complete results of the hard function.

\section{Numerical results }
\label{sec:num}

We parameterize the phase space of $gb\to tW$ process using the dimensionless variables $\beta_t$ and $\cos \theta$  in the center-of-mass frame of incoming partons, which are defined by
\begin{align}
    \beta_t = |\vec{k}_4|/k_4^0, \qquad \cos\theta = \vec{k}_1\cdot \vec{k}_4/|\vec{k}_1||\vec{k}_4|\;.
\end{align}
Then the Mandelstam variables read
\begin{align}
    s=m_W^2-m_t^2+2\Delta\,, \quad
    t=m_t^2-\Delta (1+\beta_t\cos\theta)\,,     \quad  u=m_t^2-\Delta(1-\beta_t\cos\theta)\,
\end{align}
with $\Delta =  \left(m_t^2+m_t\sqrt{m_W^2+\beta_t^2(m_t^2-m_W^2)}\right) /(1-\beta_t^2)$.  
The full phase space spans over $0\le \beta_t<1$ and $-1\le \cos\theta \le 1$~. 
In practice, we have generated a grid of $80\times 42$ phase space points and computed the amplitude squared on this grid\footnote{The numerical result on this grid is available upon request from the authors. }.
For simplicity, we extracted a factor $e^2 g_s^2/\sin^2{\theta_W}$ in the numerical results of the hard functions shown in the figures.
In numerical calculation, the $W$ boson mass is set by a rational identity $m_W^2/m_t^2=3/14$, which can significantly speed up the IBP reduction procedure.
The renormalization scale is chosen to $\mu= m_t$.

\begin{figure}[ht]
    \centering
    \includegraphics[width=0.49\textwidth]{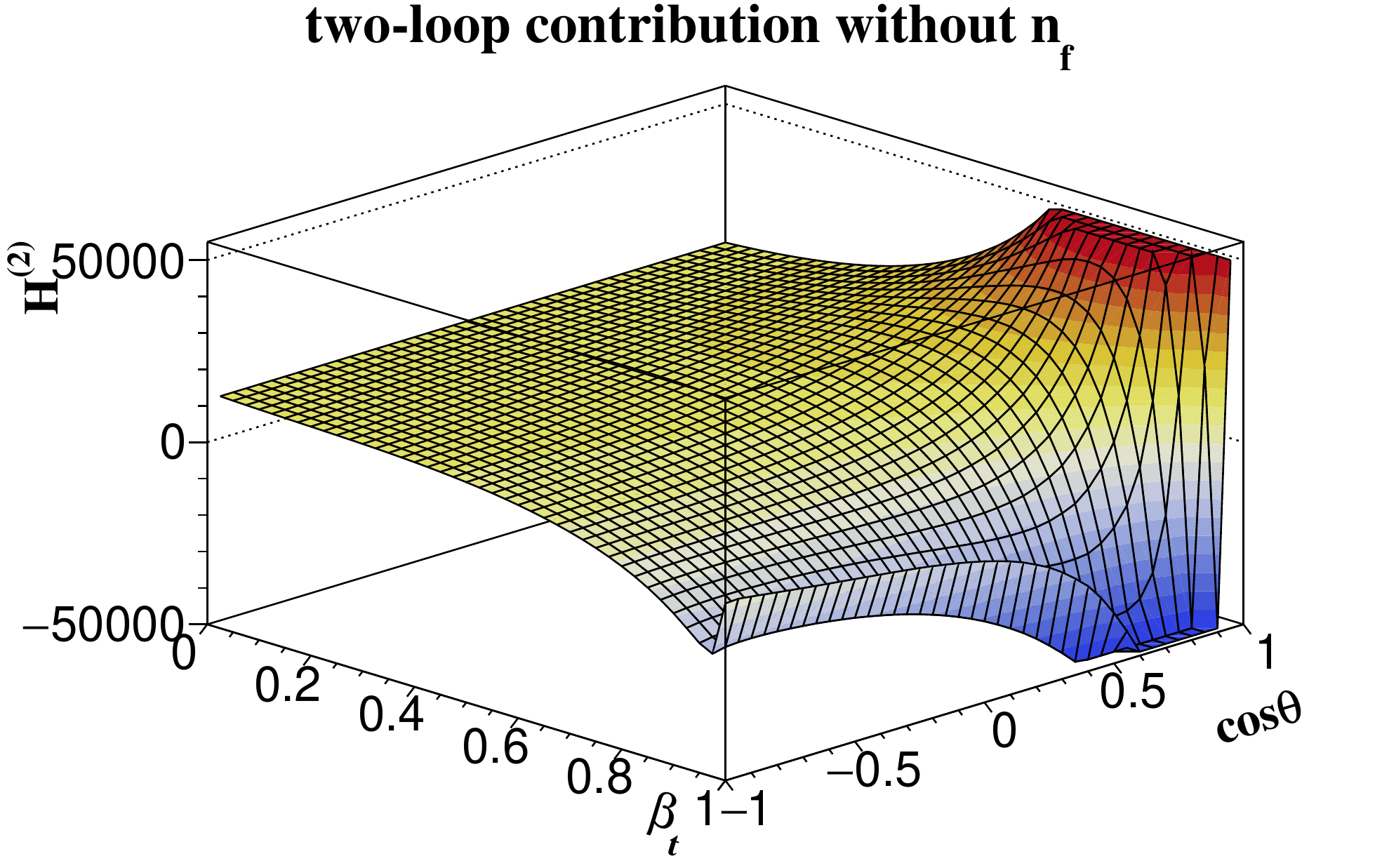}
    \includegraphics[width=0.49\textwidth]{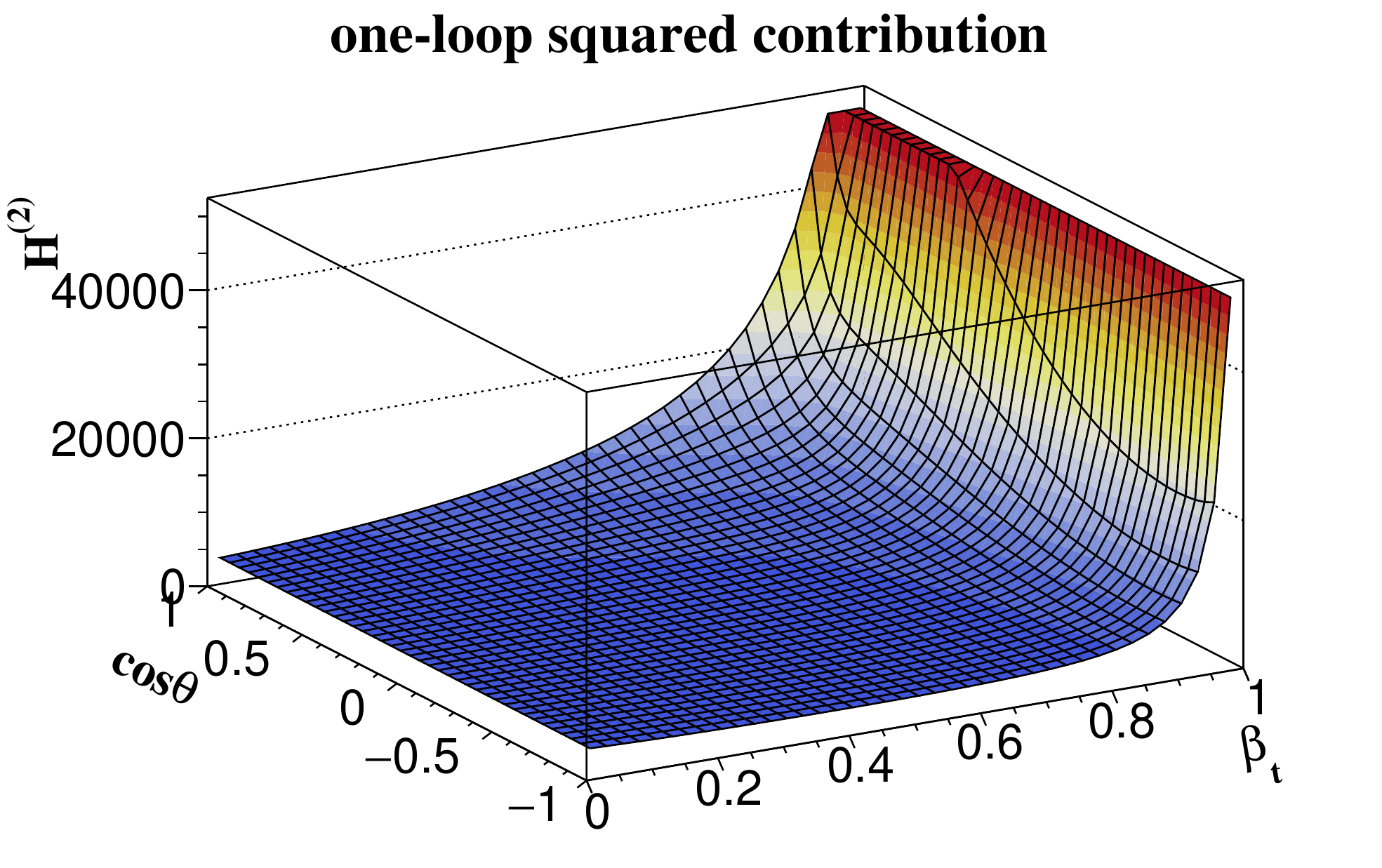}
    \includegraphics[width=0.49\textwidth]{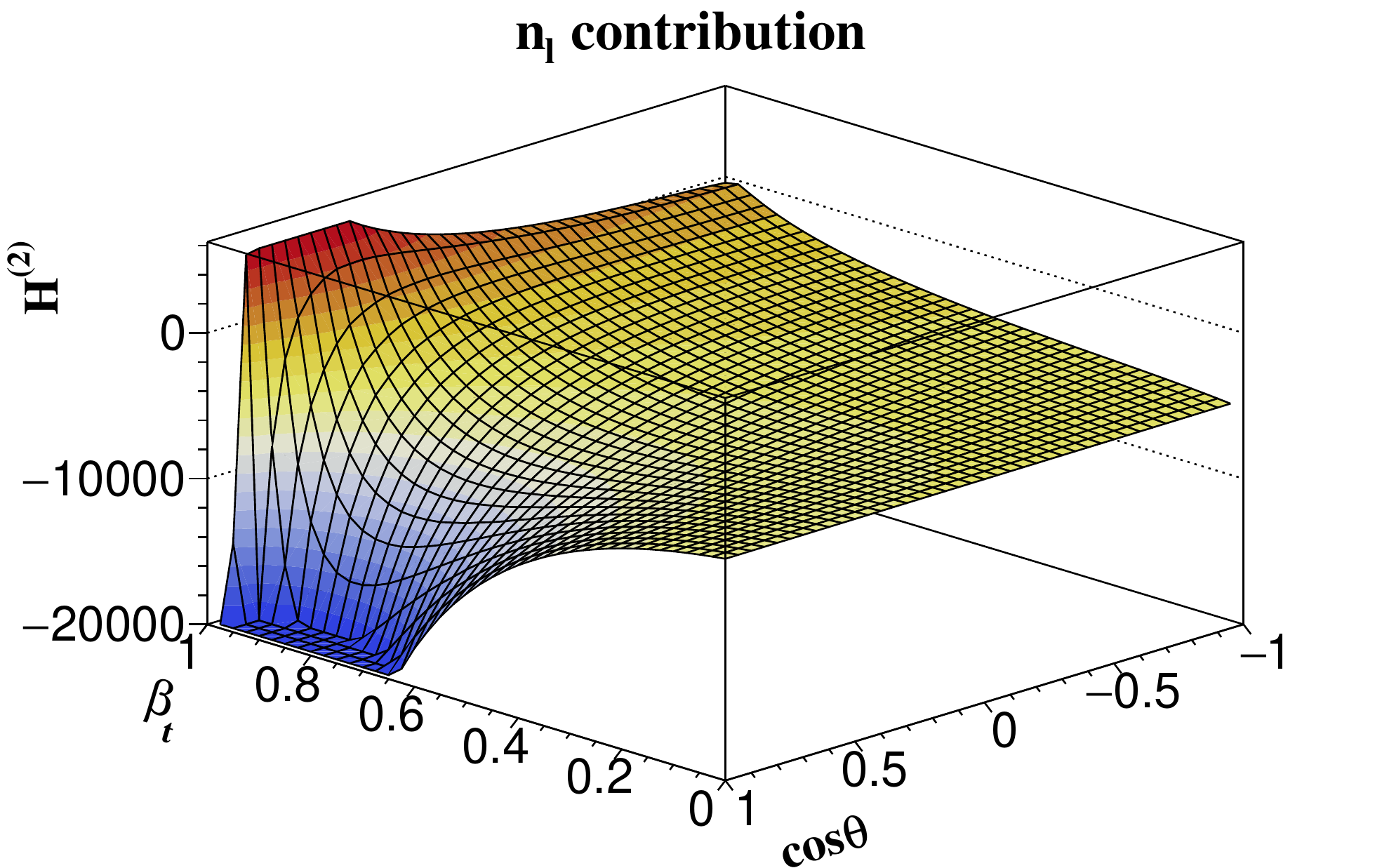}
    \includegraphics[width=0.49\textwidth]{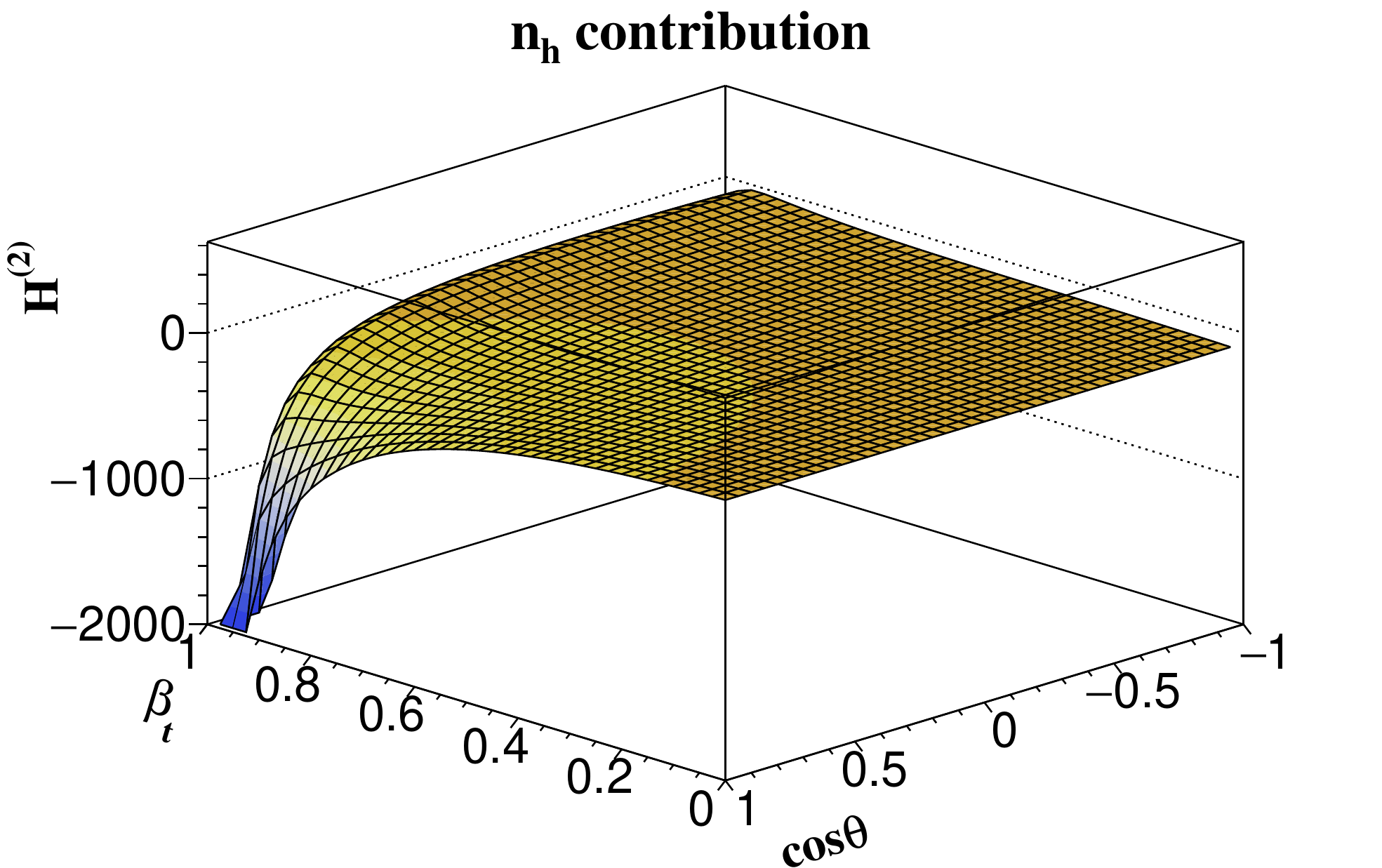}
    \caption{Different contributions to the NNLO hard function as a function of $\beta_t$ and $\cos\theta$.  Note that the 3-D diagrams have been rotated to an appropriate angle for visualization convenience. The $n_l$ ($n_h$) contribution denotes the light (heavy) quark loop correction. }
    \label{fig:h2}
\end{figure}

\begin{figure}[ht]
    \centering
    \includegraphics[width=0.7\textwidth]{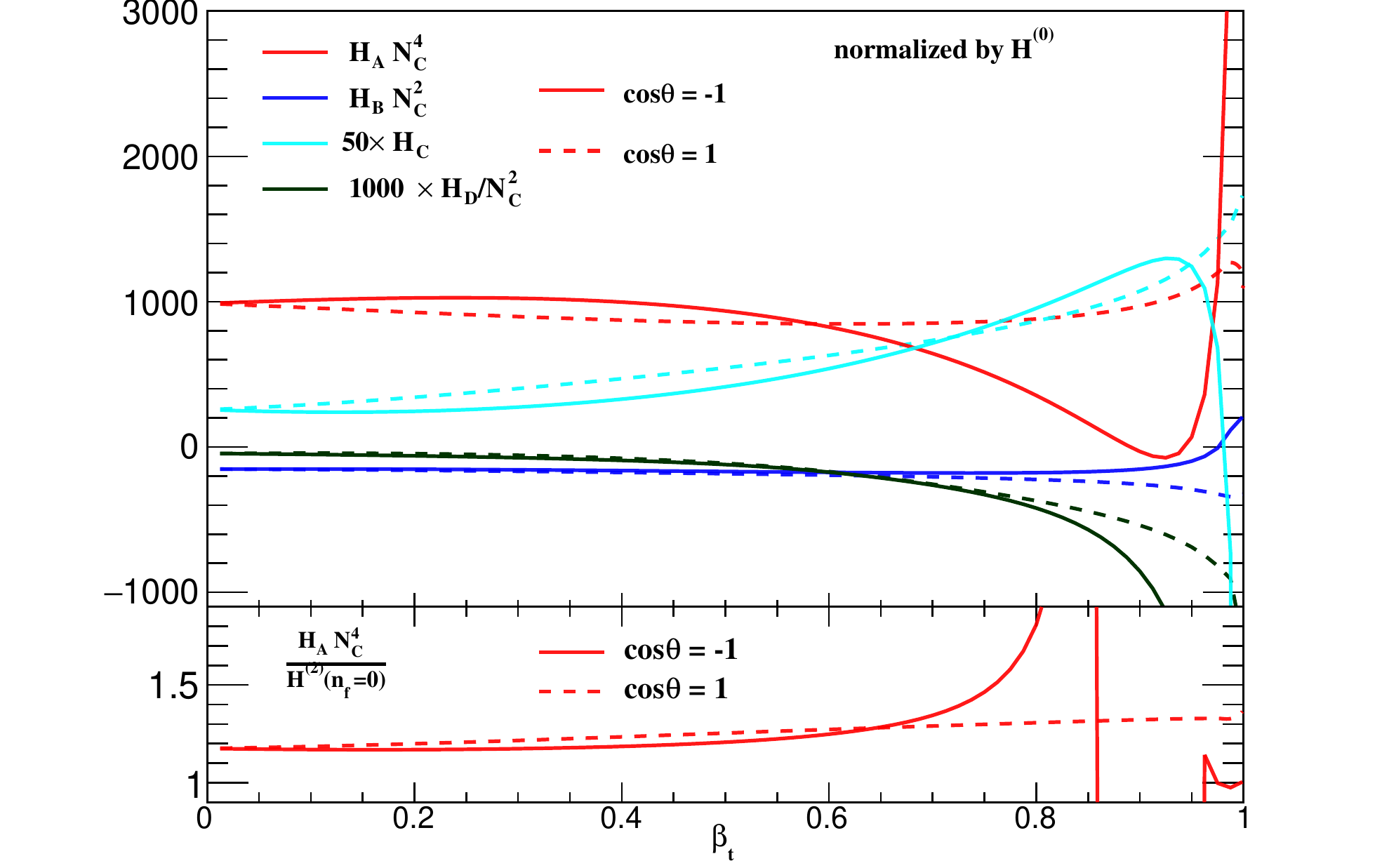}
    \caption{Contributions of different color structures in the NNLO hard function without quark loops normalized by the LO result. The red, blue, cyan, and dark blue lines correspond to $H_A N_C^4$, $H_B N_C^2$, $50\times H_C$ and $1000\times H_D/N_C^2$, respectively. The solid and dashed lines represent the results of $\cos\theta=-1$ and $\cos\theta=1$, respectively. The lower panel  shows the ratio of the leading color contribution to the hard function without quark loops. }
    \label{fig:colors}
\end{figure}

\begin{figure}[ht]
    \centering
    \includegraphics[width=0.7\textwidth]{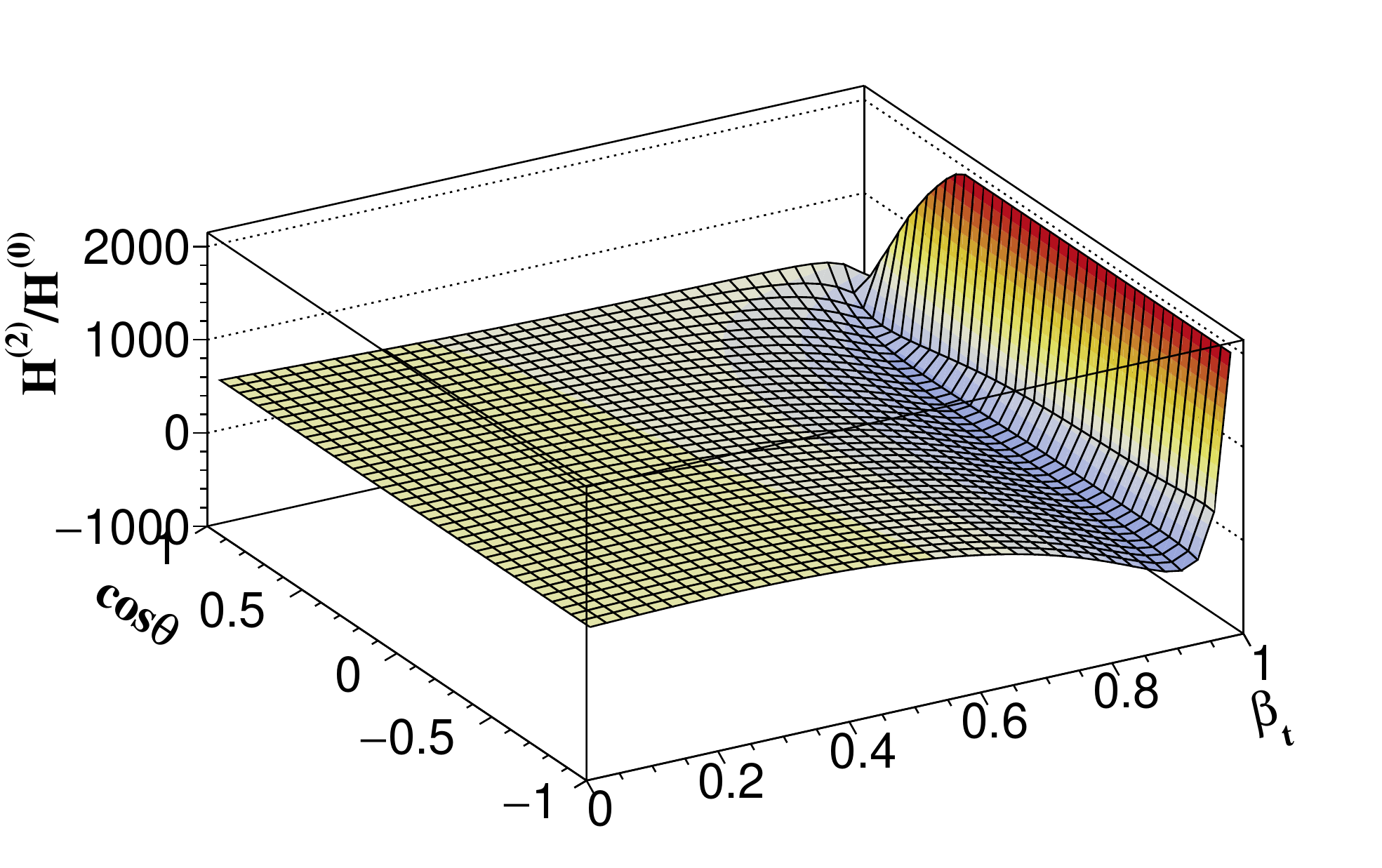}
    \caption{NNLO hard function normalized by the LO result  as a function of $\beta_t$ and $\cos\theta$.
    }
    \label{fig:h2ratio}
\end{figure}

To illustrate the size of different contributions to the hard function, we show in figure~\ref{fig:h2} the results of the two-loop corrections without quark loops, the one-loop squared, the light and heavy quark loop corrections separately in 3-D diagrams as a function of $\beta_t$ and $\cos\theta$.
One can see that these corrections are stable in a large region of $\beta_t$ and $\cos\theta$.
The two-loop  without quark loops and the one-loop squared corrections are positive while the quark-loop corrections are negative. 
However, they vary dramatically toward the boundary of $\beta_t = 1$ or $\cos\theta=1$. 
Specifically, the  heavy quark-loop contributions become negative rapidly as $\beta_t\to 1$ and $\cos\theta \to 1$, while the one-loop squared corrections are positively divergent as $\beta_t \to 1$.
The two-loop corrections without quark loops first decrease (increase) and then increase (decrease) as $\beta_t\to 1$ for $-1\le \cos\theta	\lesssim 0.5 $ ($0.5 \lesssim \cos\theta	\le 1  $).
The light quark-loop contributions drop quickly when $\cos\theta$ approaches 1 and 
$\beta_t$ is larger than about 0.6 but less than 1.  

These divergence behaviors can be understood from the structure of the squared amplitude.
The LO squared amplitude contains $1/(1-\beta_t\cos\theta)$.
This fact explains the divergence near the corner of $\beta_t=1$ and $\cos\theta=1$. 
The higher-order corrections develop additional collinear divergences in the form of 
$\ln( 1 \pm \beta_t\cos\theta)$ and $\ln( 1-\beta_t^2)$. 
The latter is responsible for the divergence away from the region with $\cos\theta=1$.
The explicit form of these logarithms can be predicted and resummed to all orders of the strong coupling following the method in Ref.~\cite{Ferroglia:2013awa}.
We leave this to future work.
For practical phenomenology study at the LHC, the suppression of the parton distribution functions near the region of $\beta_t\to 1$ dominates over these large logarithms \cite{Chen:2022ntw} and therefore it is safe to integrate over the full phase space in the calculation of the hadronic cross section.

In our previous work  \cite{Chen:2022yni}, we have provided the result of the leading color contribution.
Based on the smallness of the $1/N_C^2$ expansion parameter, we expected that it already gives the main correction.
To estimate this approximation, we show the contributions of different color structures in the NNLO hard function without quark loops ($n_f=n_l+n_h=0$) in figure~\ref{fig:colors}.
This is a 2-D diagram so that one can recognize the magnitude more clearly. 
We have drawn the curves corresponding to the parameter $\cos\theta=\pm 1$ and the results at other values of $\cos\theta$ can be considered between them.
Apart from the region around $\beta_t>0.7$, the leading color contribution is larger than the subleading color ones almost by an order of magnitude. For example 
\begin{align}
   \frac{ H^{(2)}}{H^{(0)}}\Bigg|_{\beta_t=0.4,\;\cos\theta=-1,\;n_f=0}=\underbrace{ 997}_{H_A N_C^4}  \underbrace{-162}_{H_B N_C^2} \underbrace{-6.60}_{H_C}\underbrace{-0.098}_{H_D/N_C^2}\,,
\end{align}
where the numbers on the right hand side are aligned according to the power of $N_C^2$.
From figure~\ref{fig:colors}, the divergence behaviour of the hard function near $\beta_t=1$ is also observed unambiguously.

Lastly, but most importantly, we show the numerical result of the full NNLO hard function in figure~\ref{fig:h2ratio}.
It is interesting to find that this function is flat over the large region of $\beta_t<0.8$ as a consequence of strong cancellation among different contributions.
When $\beta_t$ becomes larger than 0.8, the NNLO hard function first decreases and then increases dramatically. 
It drops down only at the corner of both $\beta_t\to 1$ and $\cos\theta\to 1$.

Notice that the hard function is not renormalization scale independent.
The scale-dependent terms can be recovered by the equation 
\begin{align}
\frac{d\ln H}{ d\ln \mu}  
    = - \frac{d  \ln {\bf  Z^* Z}} {d\ln \mu},
\end{align}
where the right hand side is easy to obtain \cite{Chen:2022ntw,Chen:2022yni}.

To have a rough estimate of the correction to the hadronic cross section at the 13 TeV LHC, we integrate over the full phase space and 
perform convolution with the CT14nlo parton distribution function \cite{Dulat:2015mca} via the interface of LHAPDF~\cite{Buckley:2014ana}.
The factorization and renormalization scales are set at $m_t$.
We find that the full NNLO hard function provides a correction of about $3\%$ to the LO cross section.

\section{Conclusion}
\label{sec:concl}
 In this paper, we present the calculation of complete two-loop amplitudes for hadronic $tW$  production. 
 The  master integrals  contain up to four massive propagators, and the corresponding differential equations involve multiple square roots that can not be rationalized simultaneously. 
 Moreover, some of the master integrals rely on several elliptic curves, which poses a challenge to analytical calculation.
We  first choose such a good integral basis using only rational transformation that the dimensional parameter is decoupled from kinematic variables in the denominators of the coefficients in the differential equations as well as the reduction coefficients in the amplitudes. 
And then we calculate the boundary conditions and solve the differential equations 
 we derived numerically with the {\tt AMFlow} package.
We have implemented a contour in the complex plane of the integration variable to maintain numerical stability in the case of pseudo poles.
We have made nontrivial checks on our calculation for both the master integrals and the full amplitudes.
In particular, the total divergences arising from many Feynman diagrams agree with the ones predicted from universal anomalous dimensions.
The finite remainder contributes to the hard function that would be used in a NNLO computation. 
The NNLO hard function is stable when the top-quark velocity $\beta_t$ is less than 0.8.
As $\beta_t$ increases, the hard function changes dramatically due to the large logarithmic enhancement from $\ln (1-\beta_t)$. 
After phase space integration and convolution with the parton distribution function, the NNLO hard function increases the LO cross section by about $3\%$.

\appendix

\section{Master integrals in the NP5, NP6 and NP7 topologies}

The 70 master integrals in the NP5 topology are chosen as
\begin{align}
\big\{& \text{I}_{2,0,0,0,2,0,0,0,0}^{\text{NP5}},~\text{I}_{0,0,2,1,2,0,0,0,0}^{\text{NP5}},\text{I}_{2,0,0,1,2,0,0,0,0}^{\text{NP5}},\text{I}_{0,0,0,1,2,2,0,0,0}^{\text{NP5}},\text{I}_{0,1,0,2,2,0,0,0,0}^{\text{NP5}},\text{I}_{0,2,0,2,1,0,0,0,0}^{\text{NP5}},\nn\\
&\text{I}_{2,0,0,0,0,2,1,0,0}^{\text{NP5}},\text{I}_{1,0,0,0,0,2,2,0,0}^{\text{NP5}},\text{I}_{2,0,0,0,2,1,0,0,0}^{\text{NP5}},\text{I}_{2,0,0,0,1,2,0,0,0}^{\text{NP5}},\text{I}_{2,2,0,0,0,0,1,0,0}^{\text{NP5}},\text{I}_{1,2,0,0,0,0,2,0,0}^{\text{NP5}},\nn\\
&\text{I}_{1,0,1,1,2,0,0,0,0}^{\text{NP5}},\text{I}_{1,0,1,0,2,1,0,0,0}^{\text{NP5}},\text{I}_{0,2,1,1,0,0,1,0,0}^{\text{NP5}},\text{I}_{1,2,0,1,0,0,1,0,0}^{\text{NP5}},\text{I}_{2,0,0,0,1,1,1,0,0}^{\text{NP5}},\text{I}_{3,0,0,0,1,1,1,0,0}^{\text{NP5}}, \nn\\
& \text{I}_{2,1,0,0,1,0,1,0,0}^{\text{NP5}},\text{I}_{3,1,0,0,1,0,1,0,0}^{\text{NP5}},\text{I}_{0,0,2,0,1,1,1,0,0}^{\text{NP5}},\text{I}_{0,0,3,0,1,1,1,0,0}^{\text{NP5}},\text{I}_{0,0,1,1,2,1,0,0,0}^{\text{NP5}},\text{I}_{0,0,1,1,3,1,0,0,0}^{\text{NP5}},\nn\\
& \text{I}_{0,0,2,1,2,1,0,0,0}^{\text{NP5}},\text{I}_{1,0,0,1,2,1,0,0,0}^{\text{NP5}},\text{I}_{1,0,0,1,3,1,0,0,0}^{\text{NP5}},\text{I}_{2,0,0,1,2,1,0,0,0}^{\text{NP5}},\text{I}_{1,1,0,1,2,0,0,0,0}^{\text{NP5}},\text{I}_{1,1,0,1,3,0,0,0,0}^{\text{NP5}},\nn\\
& \text{I}_{2,1,0,1,2,0,0,0,0}^{\text{NP5}},\text{I}_{1,1,1,0,1,0,1,0,0}^{\text{NP5}},\text{I}_{0,1,1,1,1,0,1,0,0}^{\text{NP5}},\text{I}_{0,1,2,1,1,0,1,0,0}^{\text{NP5}},\text{I}_{1,0,1,0,1,1,1,0,0}^{\text{NP5}},\text{I}_{1,0,1,0,1,2,1,0,0}^{\text{NP5}},\nn\\
& \text{I}_{1,-1,1,0,1,2,1,0,0}^{\text{NP5}},\text{I}_{1,0,1,0,2,1,1,0,0}^{\text{NP5}},\text{I}_{1,0,1,1,1,1,0,0,0}^{\text{NP5}},\text{I}_{1,0,1,1,1,2,0,0,0}^{\text{NP5}},\text{I}_{1,0,1,1,2,1,0,0,0}^{\text{NP5}},\text{I}_{1,1,0,0,1,1,1,0,0}^{\text{NP5}},\nn\\
&\text{I}_{1,1,0,0,2,1,1,0,0}^{\text{NP5}},\text{I}_{2,1,0,0,1,1,1,0,0}^{\text{NP5}},\text{I}_{1,1,0,1,1,1,0,0,0}^{\text{NP5}},\text{I}_{2,1,0,1,1,1,0,0,0}^{\text{NP5}},\text{I}_{1,1,0,1,2,1,0,0,0}^{\text{NP5}},\text{I}_{1,1,1,0,0,1,1,0,0}^{\text{NP5}},\nn\\
&\text{I}_{1,1,1,0,0,1,2,0,0}^{\text{NP5}},\text{I}_{1,1,1,1,0,0,1,0,0}^{\text{NP5}},\text{I}_{1,1,1,1,0,0,2,0,0}^{\text{NP5}},\text{I}_{1,1,0,1,1,0,1,0,0}^{\text{NP5}},\text{I}_{1,1,0,1,2,0,1,0,0}^{\text{NP5}},\text{I}_{2,1,0,1,1,0,1,0,0}^{\text{NP5}},\nn\\
&\text{I}_{1,2,0,1,1,0,1,0,0}^{\text{NP5}},\text{I}_{1,1,0,2,1,0,1,0,0}^{\text{NP5}},\text{I}_{0,1,1,1,1,1,1,0,0}^{\text{NP5}},\text{I}_{1,1,0,1,1,1,1,0,0}^{\text{NP5}},\text{I}_{1,1,0,1,1,1,1,-1,0}^{\text{NP5}},\text{I}_{1,1,-1,1,1,1,1,0,0}^{\text{NP5}},\nn\\
&\text{I}_{1,1,1,0,1,1,1,0,0}^{\text{NP5}},\text{I}_{1,1,1,0,1,1,1,0,-1}^{\text{NP5}},\text{I}_{1,1,1,0,1,1,1,-1,0}^{\text{NP5}},\text{I}_{1,1,1,1,1,0,1,0,0}^{\text{NP5}},\text{I}_{1,1,1,1,1, 0,1,-1,0}^{\text{NP5}},\text{I}_{1,1,1,1,1,0,1,0,-1}^{\text{NP5}},\nn\\
&\text{I}_{1,1,1,1,1,1,1,0,0}^{\text{NP5}},\text{I}_{1,1,1,1,1,1,1,-1,0}^{\text{NP5}},\text{I}_{1,1,1,1,1,1,1,0,-1}^{\text{NP5}},\text{I}_{1,1,1,1,1,1,1,-1,-1}^{\text{NP5}} \big\} .
\end{align}

The 80 master integrals in the NP6 topology are chosen as
\begin{align}
\big\{&\text{I}_{2,0,0,0,2,0,0,0,0}^{\text{NP6}},\text{I}_{2,0,0,0,2,0,1,0,0}^{\text{NP6}},\text{I}_{0,0,2,1,2,0,0,0,0}^{\text{NP6}},\text{I}_{2,0,0,0,0,2,1,0,0}^{\text{NP6}},\text{I}_{0,0,0,1,2,2,0,0,0}^{\text{NP6}},\text{I}_{0,0,0,2,1,2,0,0,0}^{\text{NP6}},\nn\\
&\text{I}_{0,1,0,2,2,0,0,0,0}^{\text{NP6}},\text{I}_{0,2,0,2,1,0,0,0,0}^{\text{NP6}},\text{I}_{2,0,0,0,2,1,0,0,0}^{\text{NP6}},\text{I}_{2,0,0,0,1,2,0,0,0}^{\text{NP6}},\text{I}_{2,1,0,0,0,0,2,0,0}^{\text{NP6}},\text{I}_{1,2,0,0,0,0,2,0,0}^{\text{NP6}},\nn\\
&\text{I}_{0,0,2,1,2,0,1,0,0}^{\text{NP6}},\text{I}_{1,0,1,1,2,0,0,0,0}^{\text{NP6}},\text{I}_{0,2,0,1,1,0,1,0,0}^{\text{NP6}},\text{I}_{0,2,1,1,0,0,1,0,0}^{\text{NP6}},\text{I}_{0,0,2,0,1,1,1,0,0}^{\text{NP6}},\text{I}_{0,0,3,0,1,1,1,0,0}^{\text{NP6}},\nn\\
&\text{I}_{0,0,2,0,2,1,1,0,0}^{\text{NP6}},\text{I}_{0,0,1,1,2,1,0,0,0}^{\text{NP6}},\text{I}_{0,0,1,1,3,1,0,0,0}^{\text{NP6}},\text{I}_{0,0,2,1,2,1,0,0,0}^{\text{NP6}},\text{I}_{2,0,0,0,1,1,1,0,0}^{\text{NP6}},\text{I}_{3,0,0,0,1,1,1,0,0}^{\text{NP6}},
 \nn\\
& \text{I}_{2,0,0,0,2,1,1,0,0}^{\text{NP6}},\text{I}_{1,0,0,1,2,1,0,0,0}^{\text{NP6}},\text{I}_{1,0,0,1,3,1,0,0,0}^{\text{NP6}},\text{I}_{1,0,1,0,2,1,0,0,0}^{\text{NP6}},\text{I}_{2,1,0,0,1,0,1,0,0}^{\text{NP6}},\text{I}_{3,1,0,0,1,0,1,0,0}^{\text{NP6}},\nn\\
&\text{I}_{2,1,0,0,2,0,1,0,0}^{\text{NP6}},\text{I}_{1,1,0,1,2,0,0,0,0}^{\text{NP6}},\text{I}_{1,1,0,1,3,0,0,0,0}^{\text{NP6}},\text{I}_{1,0,1,1,2,0,1,0,0}^{\text{NP6}},\text{I}_{0,0,1,1,1,1,1,0,0}^{\text{NP6}},\text{I}_{1,1,1,1,0,0,1,0,0}^{\text{NP6}},\nn\\
&\text{I}_{1,1,1,1,0,0,2,0,0}^{\text{NP6}},\text{I}_{1,0,1,1,1,1,0,0,0}^{\text{NP6}},\text{I}_{1,0,1,1,1,2,0,0,0}^{\text{NP6}},\text{I}_{1,0,1,1,2,1,0,0,0}^{ \text{NP6}},\text{I}_{0,1,1,1,1,0,1,0,0}^{\text{NP6}},\text{I}_{0,1,2,1,1,0,1,0,0}^{\text{NP6}},\nn\\
&\text{I}_{0,1,1,1,2,0,1,0,0}^{\text{NP6}},\text{I}_{1,1,0,0,1,1,1,0,0}^{\text{NP6}},\text{I}_{1,1,0,0,2,1,1,0,0}^{\text{NP6}},\text{I}_{2,1,0,0,1,1,1,0,0}^{\text{NP6}},\text{I}_{1,1,0,1,1,1,0,0,0}^{\text{NP6}},\text{I}_{2,1,0,1,1,1,0,0,0}^{\text{NP6}},\nn\\
&\text{I}_{1,1,0,1,2,1,0,0,0}^{\text{NP6}},\text{I}_{1,1,1,0,0,1,1,0,0}^{\text{NP6}},\text{I}_{1,1,1,0,0,1,2,0,0}^{\text{NP6}},\text{I}_{1,1,1,0,1,0,1,0,0}^{\text{NP6}},\text{I}_{1,1,2,0,1,0,1,0,0}^{\text{NP6}},\text{I}_{1,0,1,0,1,1,1,0,0}^{\text{NP6}},\nn\\
& \text{I}_{1,0,1,0,2,1,1,0,0}^{\text{NP6}},\text{I}_{1,0,1,0,1,2,1,0,0}^{\text{NP6}},\text{I}_{1,0,1,0,1,2,1,-1,0}^{\text{NP6}},\text{I}_{1,0,1,0,1,2,1,0,-1}^{\text{NP6}},\text{I}_{1,1,0,1,1,0,1,0,0}^{\text{NP6}},\text{I}_{2,1,0,1,1,0,1,0,0}^{\text{NP6}},\nn\\
&\text{I}_{1,1,0,1,2,0,1,0,0}^{\text{NP6}},\text{I}_{1,2,0,1,1,0,1,0,0}^{\text{NP6}},\text{I}_{1,1,0,1,1,0,2,0,0}^{\text{NP6}},\text{I}_{0,1,1,1,1,1,1,0,0}^{\text{NP6}},\text{I}_{1,0,1,1,1,1,1,0,0}^{\text{NP6}},\text{I}_{1,1,0,1,1,1,1,0,0}^{\text{NP6}},\nn\\
&\text{I}_{1,1,0,1,1,1,1,-1,0}^{\text{NP6}},\text{I}_{1,1,0,1,1,1,1,0,-1}^{\text{NP6}},\text{I}_{1,1,1,0,1,1,1,0,0}^{\text{NP6}},\text{I}_{1,1,1,0,1,1,1,-1,0}^{\text{NP6}},\text{I}_{1,1,1,0,1,1,1,0,-1}^{\text{NP6}},\text{I}_{1,1,1,0,1,1,2,-1,0}^{\text{NP6}},\nn\\
&\text{I}_{1,1,1,1,1,0,1,0,0}^{\text{NP6}},\text{I}_{1,1,1,1,1,0,1,-1,0}^{\text{NP6}},\text{I}_{1,1,1,1,1,0,1,0,-1}^{\text{NP6}},\text{I}_{1,1,1,1,1,-1,1,0,0}^{\text{NP6}},\text{I}_{1,1,1,1,1,1,1,0,0}^{\text{NP6}},\text{I}_{1,1,1,1,1,1,1,-1,0}^{\text{NP6}},\nn\\
&\text{I}_{1,1,1,1,1,1,1,0,-1}^{\text{NP6}},\text{I}_{1,1,1,1,1,1,1,-1,-1}^{\text{NP6}} \big\}.
\end{align}

The 90 master integrals in the NP7 topology are chosen as
\begin{align}
\big\{ & \text{I}^{\text{NP7}}_{2, 2, 0, 0, 0, 0, 0, 0, 0}, \text{I}^{\text{NP7}}_{0, 0, 2, 2, 0, 1, 0, 0, 0}, \text{I}^{\text{NP7}}_{2, 2, 0, 0, 0, 0, 1, 0, 0},\text{I}^{\text{NP7}}_{2, 2, 0, 0, 0, 1, 0, 0, 0},\text{I}^{\text{NP7}}_{2, 2, 0, 0, 1, 0, 0, 0, 0}, \text{I}^{\text{NP7}}_{0, 0, 2, 2, 0, 0, 1, 0, 0},  \nn\\
& \text{I}^{\text{NP7}}_{0, 0, 2, 1, 0, 0, 2, 0, 0}, \text{I}^{\text{NP7}}_{0, 1, 0, 1, 1, 0, 0, 0, 0},\text{I}^{\text{NP7}}_{0, 2, 0, 1, 1, 0, 0, 0, 0}, \text{I}^{\text{NP7}}_{2, 0, 0, 0, 2, 0, 1, 0, 0}, \text{I}^{\text{NP7}}_{1, 0, 0, 0, 2, 0, 2, 0, 0}, \text{I}^{\text{NP7}}_{2, 0, 2, 0, 0, 0, 1, 0, 0},\nn\\
& \text{I}^{\text{NP7}}_{1, 0, 2, 0, 0, 0, 2, 0, 0}, \text{I}^{\text{NP7}}_{2, 0, 2, 0, 0, 1, 0, 0, 0}, \text{I}^{\text{NP7}}_{1, 0, 2, 0, 0, 2, 0, 0, 0}, \text{I}^{\text{NP7}}_{1, 0, 0, 1, 2, 0, 1, 0, 0}, \text{I}^{\text{NP7}}_{0, 1, 1, 2, 1, 0, 0, 0, 0}, \text{I}^{\text{NP7}}_{0, 1, 0, 1, 2, 0, 1, 0, 0}, \nn\\
& \text{I}^{\text{NP7}}_{0, 2, 0, 1, 2, 0, 1, 0, 0}, \text{I}^{\text{NP7}}_{0, 0, 1, 2, 1, 0, 1, 0, 0}, \text{I}^{\text{NP7}}_{0, 0, 1, 3, 1, 0, 1, 0, 0}, \text{I}^{\text{NP7}}_{0, 1, 1, 2, 0, 0, 1, 0, 0}, \text{I}^{\text{NP7}}_{0, 1, 1, 3, 0, 0, 1, 0, 0}, \text{I}^{\text{NP7}}_{0, 2, 1, 2, 0, 0, 1, 0, 0}, \nn\\
& \text{I}^{\text{NP7}}_{0, 1, 1, 2, 0, 1, 0, 0, 0}, \text{I}^{\text{NP7}}_{0, 1, 1, 3, 0, 1, 0, 0, 0}, \text{I}^{\text{NP7}}_{0, 2, 1, 2, 0, 1, 0, 0, 0},  \text{I}^{\text{NP7}}_{2, 0, 1, 0, 1, 0, 1, 0, 0},  \text{I}^{\text{NP7}}_{3, 0, 1, 0, 1, 0, 1, 0, 0}, \text{I}^{\text{NP7}}_{2, 0, 1, 0, 1, 1, 0, 0, 0}, \nn\\
& \text{I}^{\text{NP7}}_{3, 0, 1, 0, 1, 1, 0, 0, 0}, \text{I}^{\text{NP7}}_{1, 1, 0, 0, 2, 0, 1, 0, 0},  \text{I}^{\text{NP7}}_{1, 1, 0, 0, 2, 0, 2, 0, 0}, \text{I}^{\text{NP7}}_{2, 1, 0, 0, 2, 0, 1, 0, 0}, \text{I}^{\text{NP7}}_{1, 1, 0, 0, 2, 1, 0, 0, 0}, \text{I}^{\text{NP7}}_{1, 2, 0, 0, 2, 1, 0, 0, 0}, \nn\\
& \text{I}^{\text{NP7}}_{1, 1, 0, 1, 2, 0, 0, 0, 0}, \text{I}^{\text{NP7}}_{1, 0, 1, 1, 1, 1, 0, 0, 0}, \text{I}^{\text{NP7}}_{1, 1, 0, 1, 1, 1, 0, 0, 0},  \text{I}^{\text{NP7}}_{1, 1, 1, 1, 0, 0, 1, 0, 0}, \text{I}^{\text{NP7}}_{1, 1, 1, 1, 0, 0, 2, 0, 0}, \text{I}^{\text{NP7}}_{1, 1, 1, 1, 0, 1, 0, 0, 0}, \nn\\
&\text{I}^{\text{NP7}}_{1, 1, 1, 1, 0, 2, 0, 0, 0}, \text{I}^{\text{NP7}}_{0, 1, 1, 1, 0, 1, 1, 0, 0}, \text{I}^{\text{NP7}}_{0, 1, 1, 2, 0, 1, 1, 0, 0}, \text{I}^{\text{NP7}}_{0, 1, 1, 1, 1, 1, 0, 0, 0}, \text{I}^{\text{NP7}}_{0, 1, 1, 2, 1, 1, 0, 0, 0}, \text{I}^{\text{NP7}}_{1, 0, 1, 0, 1, 1, 1, 0, 0}, \nn\\
& \text{I}^{\text{NP7}}_{1, 0, 1, 0, 2, 1, 1, 0, 0}, \text{I}^{\text{NP7}}_{2, 0, 1, 0, 1, 1, 1, 0, 0}, \text{I}^{\text{NP7}}_{1, 0, 1, 1, 0, 1, 1, 0, 0}, \text{I}^{\text{NP7}}_{2, 0, 1, 1, 0, 1, 1, 0, 0},  \text{I}^{\text{NP7}}_{1, 0, 1, 1, 1, 0, 1, 0, 0}, \text{I}^{\text{NP7}}_{1, 0, 1, 1, 2, 0, 1, 0, 0}, \nn\\
& \text{I}^{\text{NP7}}_{1, 0, 1, 1, 1, 0, 2, 0, 0}, \text{I}^{\text{NP7}}_{1, 0, 1, 1, 1, -1, 2, 0, 0},  \text{I}^{\text{NP7}}_{1, 1, 0, 0, 1, 1, 1, 0, 0}, \text{I}^{\text{NP7}}_{1, 1, 0, 0, 2, 1, 1, 0, 0}, \text{I}^{\text{NP7}}_{1, 1, 0, 1, 1, 0, 1, 0, 0}, \text{I}^{\text{NP7}}_{1, 1, 0, 1, 1, 0, 2, 0, 0}, \nn\\
& \text{I}^{\text{NP7}}_{1, 1, 0, 1, 2, 0, 1, 0, 0}, \text{I}^{\text{NP7}}_{1, 1, 0, 2, 1, 0, 1, 0, 0}, \text{I}^{\text{NP7}}_{2, 1, 0, 1, 1, 0, 1, 0, 0},  \text{I}^{\text{NP7}}_{1, 2, 0, 1, 1, 0, 1, 0, 0},  \text{I}^{\text{NP7}}_{0, 1, 1, 1, 1, 0, 1, 0, 0}, \text{I}^{\text{NP7}}_{0, 1, 1, 1, 1, 0, 2, 0, 0}, \nn\\
&\text{I}^{\text{NP7}}_{0, 1, 1, 1, 2, 0, 1, 0, 0}, \text{I}^{\text{NP7}}_{0, 1, 1, 2, 1, 0, 1, 0, 0},  \text{I}^{\text{NP7}}_{0, 1, 2, 1, 1, 0, 1, 0, 0}, \text{I}^{\text{NP7}}_{0, 2, 1, 1, 1, 0, 1, 0, 0}, \text{I}^{\text{NP7}}_{0, 1, 1, 1, 1, 0, 2, 0, -1}, \text{I}^{\text{NP7}}_{1, 1, 0, 1, 1, 1, 1, 0, 0},\nn\\
& \text{I}^{\text{NP7}}_{1, 1, 1, 1, 0, 1, 1, 0, 0}, \text{I}^{\text{NP7}}_{0, 1, 1, 1, 1, 1, 1, 0, 0}, \text{I}^{\text{NP7}}_{1, 1, 1, 1, 1, 1, 0, 0, 0},  \text{I}^{\text{NP7}}_{1, 1, 1, 1, 1, 1, 0, 0, -1}, \text{I}^{\text{NP7}}_{1, 0, 1, 1, 1, 1, 1, 0, 0}, \text{I}^{\text{NP7}}_{1, 0, 1, 1, 1, 1, 1, -1, 0},  \nn\\
& \text{I}^{\text{NP7}}_{1, 0, 1, 1, 1, 1, 1, 0, -1}, \text{I}^{\text{NP7}}_{1, 1, 1, 1, 1, 0, 1, 0, 0}, \text{I}^{\text{NP7}}_{1, 1, 1, 1, 1, 0, 1, -1, 0}, \text{I}^{\text{NP7}}_{1, 1, 1, 1, 1, 0, 1, 0, -1}, \text{I}^{\text{NP7}}_{1, 1, 1, 1, 1, -1, 1, 0, 0}, \text{I}^{\text{NP7}}_{1, 1, 1, 2, 1, 0, 1, 0, 0}, \nn\\
& \text{I}^{\text{NP7}}_{2, 1, 1, 1, 1, 0, 1, 0, 0}, \text{I}^{\text{NP7}}_{1, 1, 1, 1, 1, 1, 1, 0, 0}, \text{I}^{\text{NP7}}_{1, 1, 1, 1, 1, 1, 1, 0, -1}, \text{I}^{\text{NP7}}_{1, 1, 1, 1, 1, 1, 1, -1, 0}, \text{I}^{\text{NP7}}_{1, 1, 1, 1, 1, 1, 1, -1, -1}, \text{I}^{\text{NP7}}_{1, 1, 1, 1, 1, 1, 1, 0, -2}\big\}.
\end{align}
The master integrals of other families are listed in the ancillary file on arXiv. The differential equations for all the families can be obtained from the authors upon request.

\section*{Acknowledgements}
This work was supported in part by the National Science Foundation of China under grants Nos. 12005117,   12075251,  12147154, 12175048,   and 12275156.  The work of L.D. and J.W. was also supported by the Taishan Scholar Foundation of Shandong province (tsqn201909011).

\bibliographystyle{JHEP}
\bibliography{2loop}

\providecommand{\href}[2]{#2}\begingroup\raggedright\begin{thebibliography}{10}

\bibitem{D0:1995jca}
{\scshape D0} collaboration, S.~Abachi et~al., \emph{{Observation of the top
  quark}}, \href{https://doi.org/10.1103/PhysRevLett.74.2632}{\emph{Phys. Rev.
  Lett.} {\bfseries 74} (1995) 2632--2637},
  [\href{https://arxiv.org/abs/hep-ex/9503003}{{\ttfamily hep-ex/9503003}}].

\bibitem{CDF:1995wbb}
{\scshape CDF} collaboration, F.~Abe et~al., \emph{{Observation of top quark
  production in $\bar{p}p$ collisions}},
  \href{https://doi.org/10.1103/PhysRevLett.74.2626}{\emph{Phys. Rev. Lett.}
  {\bfseries 74} (1995) 2626--2631},
  [\href{https://arxiv.org/abs/hep-ex/9503002}{{\ttfamily hep-ex/9503002}}].

\bibitem{ATLAS:2016ofl}
{\scshape ATLAS} collaboration, M.~Aaboud et~al., \emph{{Measurement of the
  cross-section for producing a W boson in association with a single top quark
  in pp collisions at $ \sqrt{s}=13 $ TeV with ATLAS}},
  \href{https://doi.org/10.1007/JHEP01(2018)063}{\emph{JHEP} {\bfseries 01}
  (2018) 063}, [\href{https://arxiv.org/abs/1612.07231}{{\ttfamily
  1612.07231}}].

\bibitem{ATLAS:2017quy}
{\scshape ATLAS} collaboration, M.~Aaboud et~al., \emph{{Measurement of
  differential cross-sections of a single top quark produced in association
  with a $W$ boson at $\sqrt{s}=13$ TeV with ATLAS}},
  \href{https://doi.org/10.1140/epjc/s10052-018-5649-8}{\emph{Eur. Phys. J. C}
  {\bfseries 78} (2018) 186},
  [\href{https://arxiv.org/abs/1712.01602}{{\ttfamily 1712.01602}}].

\bibitem{CMS:2021vqm}
{\scshape CMS} collaboration, A.~Tumasyan et~al., \emph{{Observation of tW
  production in the single-lepton channel in pp collisions at $ \sqrt{s} $ = 13
  TeV}}, \href{https://doi.org/10.1007/JHEP11(2021)111}{\emph{JHEP} {\bfseries
  11} (2021) 111}, [\href{https://arxiv.org/abs/2109.01706}{{\ttfamily
  2109.01706}}].

\bibitem{CMS:2022ytw}
{\scshape CMS} collaboration, \emph{{Measurement of inclusive and differential
  cross sections for single top quark production in association with a W boson
  in proton-proton collisions at $\sqrt{s}$ = 13 TeV}},
  \href{https://arxiv.org/abs/2208.00924}{{\ttfamily 2208.00924}}.

\bibitem{Giele:1995kr}
W.~T. Giele, S.~Keller and E.~Laenen, \emph{{QCD corrections to $W$ boson plus
  heavy quark production at the Tevatron}},
  \href{https://doi.org/10.1016/0370-2693(96)00078-0}{\emph{Phys. Lett.}
  {\bfseries B372} (1996) 141--149},
  [\href{https://arxiv.org/abs/hep-ph/9511449}{{\ttfamily hep-ph/9511449}}].

\bibitem{Zhu:2001hw}
S.~Zhu, \emph{{Next-to-leading order QCD corrections to bg $\to$ tW${}^-$ at
  CERN large hadron collider}},
  \href{https://doi.org/10.1016/S0370-2693(02)01952-4,
  10.1016/S0370-2693(01)01404-6}{\emph{Phys. Lett.} {\bfseries B524} (2002)
  283--288}, [\href{https://arxiv.org/abs/hep-ph/0109269}{{\ttfamily
  hep-ph/0109269}}].

\bibitem{Cao:2008af}
Q.-H. Cao, \emph{{Demonstration of One Cutoff Phase Space Slicing Method:
  Next-to-Leading Order QCD Corrections to the tW Associated Production in
  Hadron Collision}},  \href{https://arxiv.org/abs/0801.1539}{{\ttfamily
  0801.1539}}.

\bibitem{Kant:2014oha}
P.~Kant, O.~M. Kind, T.~Kintscher, T.~Lohse, T.~Martini, S.~Mölbitz et~al.,
  \emph{{HatHor for single top-quark production: Updated predictions and
  uncertainty estimates for single top-quark production in hadronic
  collisions}}, \href{https://doi.org/10.1016/j.cpc.2015.02.001}{\emph{Comput.
  Phys. Commun.} {\bfseries 191} (2015) 74--89},
  [\href{https://arxiv.org/abs/1406.4403}{{\ttfamily 1406.4403}}].

\bibitem{Campbell:2005bb}
J.~M. Campbell and F.~Tramontano, \emph{{Next-to-leading order corrections to
  Wt production and decay}},
  \href{https://doi.org/10.1016/j.nuclphysb.2005.08.015}{\emph{Nucl. Phys.}
  {\bfseries B726} (2005) 109--130},
  [\href{https://arxiv.org/abs/hep-ph/0506289}{{\ttfamily hep-ph/0506289}}].

\bibitem{Frixione:2008yi}
S.~Frixione, E.~Laenen, P.~Motylinski, B.~R. Webber and C.~D. White,
  \emph{{Single-top hadroproduction in association with a W boson}},
  \href{https://doi.org/10.1088/1126-6708/2008/07/029}{\emph{JHEP} {\bfseries
  07} (2008) 029}, [\href{https://arxiv.org/abs/0805.3067}{{\ttfamily
  0805.3067}}].

\bibitem{Re:2010bp}
E.~Re, \emph{{Single-top Wt-channel production matched with parton showers
  using the POWHEG method}},
  \href{https://doi.org/10.1140/epjc/s10052-011-1547-z}{\emph{Eur. Phys. J.}
  {\bfseries C71} (2011) 1547},
  [\href{https://arxiv.org/abs/1009.2450}{{\ttfamily 1009.2450}}].

\bibitem{Jezo:2016ujg}
T.~Ježo, J.~M. Lindert, P.~Nason, C.~Oleari and S.~Pozzorini, \emph{{An NLO+PS
  generator for $t\bar{t}$ and $Wt$ production and decay including non-resonant
  and interference effects}},
  \href{https://doi.org/10.1140/epjc/s10052-016-4538-2}{\emph{Eur. Phys. J.}
  {\bfseries C76} (2016) 691},
  [\href{https://arxiv.org/abs/1607.04538}{{\ttfamily 1607.04538}}].

\bibitem{Beccaria:2007tc}
M.~Beccaria, C.~M. Carloni~Calame, G.~Macorini, G.~Montagna, F.~Piccinini,
  F.~M. Renard et~al., \emph{{A Complete one-loop description of associated tW
  production at LHC and a search for possible genuine supersymmetric effects}},
  \href{https://doi.org/10.1140/epjc/s10052-007-0452-y}{\emph{Eur. Phys. J. C}
  {\bfseries 53} (2008) 257--265},
  [\href{https://arxiv.org/abs/0705.3101}{{\ttfamily 0705.3101}}].

\bibitem{Kidonakis:2006bu}
N.~Kidonakis, \emph{{Single top production at the Tevatron: Threshold
  resummation and finite-order soft gluon corrections}},
  \href{https://doi.org/10.1103/PhysRevD.74.114012}{\emph{Phys. Rev.}
  {\bfseries D74} (2006) 114012},
  [\href{https://arxiv.org/abs/hep-ph/0609287}{{\ttfamily hep-ph/0609287}}].

\bibitem{Kidonakis:2010ux}
N.~Kidonakis, \emph{{Two-loop soft anomalous dimensions for single top quark
  associated production with a $W^-$ or $H^-$}},
  \href{https://doi.org/10.1103/PhysRevD.82.054018}{\emph{Phys. Rev.}
  {\bfseries D82} (2010) 054018},
  [\href{https://arxiv.org/abs/1005.4451}{{\ttfamily 1005.4451}}].

\bibitem{Kidonakis:2016sjf}
N.~Kidonakis, \emph{{Soft-gluon corrections for $tW$ production at N$^3$LO}},
  \href{https://doi.org/10.1103/PhysRevD.96.034014}{\emph{Phys. Rev.}
  {\bfseries D96} (2017) 034014},
  [\href{https://arxiv.org/abs/1612.06426}{{\ttfamily 1612.06426}}].

\bibitem{Kidonakis:2021vob}
N.~Kidonakis and N.~Yamanaka, \emph{{Higher-order corrections for $tW$
  production at high-energy hadron colliders}},
  \href{https://doi.org/10.1007/JHEP05(2021)278}{\emph{JHEP} {\bfseries 05}
  (2021) 278}, [\href{https://arxiv.org/abs/2102.11300}{{\ttfamily
  2102.11300}}].

\bibitem{Li:2019dhg}
C.~S. Li, H.~T. Li, D.~Y. Shao and J.~Wang, \emph{{Momentum-space threshold
  resummation in $tW$ production at the LHC}},
  \href{https://doi.org/10.1007/JHEP06(2019)125}{\emph{JHEP} {\bfseries 06}
  (2019) 125}, [\href{https://arxiv.org/abs/1903.01646}{{\ttfamily
  1903.01646}}].

\bibitem{Li:2016tvb}
H.~T. Li and J.~Wang, \emph{{Next-to-Next-to-Leading Order $N$-Jettiness Soft
  Function for One Massive Colored Particle Production at Hadron Colliders}},
  \href{https://doi.org/10.1007/JHEP02(2017)002}{\emph{JHEP} {\bfseries 02}
  (2017) 002}, [\href{https://arxiv.org/abs/1611.02749}{{\ttfamily
  1611.02749}}].

\bibitem{Li:2018tsq}
H.~T. Li and J.~Wang, \emph{{Next-to-next-to-leading order $N$-jettiness soft
  function for $tW$ production}},
  \href{https://doi.org/10.1016/j.physletb.2018.08.019}{\emph{Phys. Lett.}
  {\bfseries B784} (2018) 397--404},
  [\href{https://arxiv.org/abs/1804.06358}{{\ttfamily 1804.06358}}].

\bibitem{Chen:2022yni}
L.-B. Chen, L.~Dong, H.~T. Li, Z.~Li, J.~Wang and Y.~Wang, \emph{{Analytic
  two-loop QCD amplitudes for tW production: Leading color and light
  fermion-loop contributions}},
  \href{https://doi.org/10.1103/PhysRevD.106.096029}{\emph{Phys. Rev. D}
  {\bfseries 106} (2022) 096029},
  [\href{https://arxiv.org/abs/2208.08786}{{\ttfamily 2208.08786}}].

\bibitem{Chen:2021gjv}
L.-B. Chen and J.~Wang, \emph{{Analytic two-loop master integrals for tW
  production at hadron colliders: I *}},
  \href{https://doi.org/10.1088/1674-1137/ac2a1e}{\emph{Chin. Phys. C}
  {\bfseries 45} (2021) 123106},
  [\href{https://arxiv.org/abs/2106.12093}{{\ttfamily 2106.12093}}].

\bibitem{Long:2021vse}
M.-M. Long, R.-Y. Zhang, W.-G. Ma, Y.~Jiang, L.~Han, Z.~Li et~al.,
  \emph{{Two-loop master integrals for the single top production associated
  with $W$ boson}},  \href{https://arxiv.org/abs/2111.14172}{{\ttfamily
  2111.14172}}.

\bibitem{Wang:2022enl}
J.~Wang and Y.~Wang, \emph{{Analytic two-loop master integrals for $tW$
  production at hadron colliders: II}},
  \href{https://arxiv.org/abs/2211.13713}{{\ttfamily 2211.13713}}.

\bibitem{Chen:2022ntw}
L.-B. Chen, L.~Dong, H.~T. Li, Z.~Li, J.~Wang and Y.~Wang, \emph{{One-loop
  squared amplitudes for hadronic tW production at next-to-next-to-leading
  order in QCD}}, \href{https://doi.org/10.1007/JHEP08(2022)211}{\emph{JHEP}
  {\bfseries 08} (2022) 211},
  [\href{https://arxiv.org/abs/2204.13500}{{\ttfamily 2204.13500}}].

\bibitem{Henn:2013pwa}
J.~M. Henn, \emph{{Multiloop integrals in dimensional regularization made
  simple}}, \href{https://doi.org/10.1103/PhysRevLett.110.251601}{\emph{Phys.
  Rev. Lett.} {\bfseries 110} (2013) 251601},
  [\href{https://arxiv.org/abs/1304.1806}{{\ttfamily 1304.1806}}].

\bibitem{Liu:2017jxz}
X.~Liu, Y.-Q. Ma and C.-Y. Wang, \emph{{A Systematic and Efficient Method to
  Compute Multi-loop Master Integrals}},
  \href{https://doi.org/10.1016/j.physletb.2018.02.026}{\emph{Phys. Lett. B}
  {\bfseries 779} (2018) 353--357},
  [\href{https://arxiv.org/abs/1711.09572}{{\ttfamily 1711.09572}}].

\bibitem{Hahn:2000kx}
T.~Hahn, \emph{{Generating Feynman diagrams and amplitudes with FeynArts 3}},
  \href{https://doi.org/10.1016/S0010-4655(01)00290-9}{\emph{Comput. Phys.
  Commun.} {\bfseries 140} (2001) 418--431},
  [\href{https://arxiv.org/abs/hep-ph/0012260}{{\ttfamily hep-ph/0012260}}].

\bibitem{Shtabovenko:2020gxv}
V.~Shtabovenko, R.~Mertig and F.~Orellana, \emph{{FeynCalc 9.3: New features
  and improvements}},
  \href{https://doi.org/10.1016/j.cpc.2020.107478}{\emph{Comput. Phys. Commun.}
  {\bfseries 256} (2020) 107478},
  [\href{https://arxiv.org/abs/2001.04407}{{\ttfamily 2001.04407}}].

\bibitem{Korner:1991sx}
J.~G. Korner, D.~Kreimer and K.~Schilcher, \emph{{A Practicable gamma(5) scheme
  in dimensional regularization}},
  \href{https://doi.org/10.1007/BF01559471}{\emph{Z. Phys. C} {\bfseries 54}
  (1992) 503--512}.

\bibitem{Smirnov:2019qkx}
A.~V. Smirnov and F.~S. Chuharev, \emph{{FIRE6: Feynman Integral REduction with
  Modular Arithmetic}},
  \href{https://doi.org/10.1016/j.cpc.2019.106877}{\emph{Comput. Phys. Commun.}
  {\bfseries 247} (2020) 106877},
  [\href{https://arxiv.org/abs/1901.07808}{{\ttfamily 1901.07808}}].

\bibitem{Kotikov:1990kg}
A.~V. Kotikov, \emph{{Differential equations method: New technique for massive
  Feynman diagrams calculation}},
  \href{https://doi.org/10.1016/0370-2693(91)90413-K}{\emph{Phys. Lett.}
  {\bfseries B254} (1991) 158--164}.

\bibitem{Kotikov:1991pm}
A.~V. Kotikov, \emph{{Differential equation method: The Calculation of N point
  Feynman diagrams}}, \href{https://doi.org/10.1016/0370-2693(91)90536-Y,
  10.1016/0370-2693(92)91582-T}{\emph{Phys. Lett.} {\bfseries B267} (1991)
  123--127}.

\bibitem{Goncharov:1998kja}
A.~B. Goncharov, \emph{{Multiple polylogarithms, cyclotomy and modular
  complexes}}, \href{https://doi.org/10.4310/MRL.1998.v5.n4.a7}{\emph{Math.
  Res. Lett.} {\bfseries 5} (1998) 497--516},
  [\href{https://arxiv.org/abs/1105.2076}{{\ttfamily 1105.2076}}].

\bibitem{Primo:2016ebd}
A.~Primo and L.~Tancredi, \emph{{On the maximal cut of Feynman integrals and
  the solution of their differential equations}},
  \href{https://doi.org/10.1016/j.nuclphysb.2016.12.021}{\emph{Nucl. Phys. B}
  {\bfseries 916} (2017) 94--116},
  [\href{https://arxiv.org/abs/1610.08397}{{\ttfamily 1610.08397}}].

\bibitem{Primo:2017ipr}
A.~Primo and L.~Tancredi, \emph{{Maximal cuts and differential equations for
  Feynman integrals. An application to the three-loop massive banana graph}},
  \href{https://doi.org/10.1016/j.nuclphysb.2017.05.018}{\emph{Nucl. Phys. B}
  {\bfseries 921} (2017) 316--356},
  [\href{https://arxiv.org/abs/1704.05465}{{\ttfamily 1704.05465}}].

\bibitem{Adams:2017tga}
L.~Adams, E.~Chaubey and S.~Weinzierl, \emph{{Simplifying Differential
  Equations for Multiscale Feynman Integrals beyond Multiple Polylogarithms}},
  \href{https://doi.org/10.1103/PhysRevLett.118.141602}{\emph{Phys. Rev. Lett.}
  {\bfseries 118} (2017) 141602},
  [\href{https://arxiv.org/abs/1702.04279}{{\ttfamily 1702.04279}}].

\bibitem{Harley:2017qut}
M.~Harley, F.~Moriello and R.~M. Schabinger, \emph{{Baikov-Lee Representations
  Of Cut Feynman Integrals}},
  \href{https://doi.org/10.1007/JHEP06(2017)049}{\emph{JHEP} {\bfseries 06}
  (2017) 049}, [\href{https://arxiv.org/abs/1705.03478}{{\ttfamily
  1705.03478}}].

\bibitem{Adams:2018bsn}
L.~Adams, E.~Chaubey and S.~Weinzierl, \emph{{Planar Double Box Integral for
  Top Pair Production with a Closed Top Loop to all orders in the Dimensional
  Regularization Parameter}},
  \href{https://doi.org/10.1103/PhysRevLett.121.142001}{\emph{Phys. Rev. Lett.}
  {\bfseries 121} (2018) 142001},
  [\href{https://arxiv.org/abs/1804.11144}{{\ttfamily 1804.11144}}].

\bibitem{Broedel:2019hyg}
J.~Broedel, C.~Duhr, F.~Dulat, B.~Penante and L.~Tancredi, \emph{{Elliptic
  polylogarithms and Feynman parameter integrals}},
  \href{https://doi.org/10.1007/JHEP05(2019)120}{\emph{JHEP} {\bfseries 05}
  (2019) 120}, [\href{https://arxiv.org/abs/1902.09971}{{\ttfamily
  1902.09971}}].

\bibitem{Broedel:2019kmn}
J.~Broedel, C.~Duhr, F.~Dulat, R.~Marzucca, B.~Penante and L.~Tancredi,
  \emph{{An analytic solution for the equal-mass banana graph}},
  \href{https://doi.org/10.1007/JHEP09(2019)112}{\emph{JHEP} {\bfseries 09}
  (2019) 112}, [\href{https://arxiv.org/abs/1907.03787}{{\ttfamily
  1907.03787}}].

\bibitem{Frellesvig:2019byn}
H.~Frellesvig, M.~Hidding, L.~Maestri, F.~Moriello and G.~Salvatori, \emph{{The
  complete set of two-loop master integrals for Higgs + jet production in
  QCD}}, \href{https://doi.org/10.1007/JHEP06(2020)093}{\emph{JHEP} {\bfseries
  06} (2020) 093}, [\href{https://arxiv.org/abs/1911.06308}{{\ttfamily
  1911.06308}}].

\bibitem{Walden:2020odh}
M.~Walden and S.~Weinzierl, \emph{{Numerical evaluation of iterated integrals
  related to elliptic Feynman integrals}},
  \href{https://doi.org/10.1016/j.cpc.2021.108020}{\emph{Comput. Phys. Commun.}
  {\bfseries 265} (2021) 108020},
  [\href{https://arxiv.org/abs/2010.05271}{{\ttfamily 2010.05271}}].

\bibitem{Muller:2022gec}
H.~M\"uller and S.~Weinzierl, \emph{{A Feynman integral depending on two
  elliptic curves}}, \href{https://doi.org/10.1007/JHEP07(2022)101}{\emph{JHEP}
  {\bfseries 07} (2022) 101},
  [\href{https://arxiv.org/abs/2205.04818}{{\ttfamily 2205.04818}}].

\bibitem{Pogel:2022yat}
S.~P\"ogel, X.~Wang and S.~Weinzierl, \emph{{The three-loop equal-mass banana
  integral in \ensuremath{\varepsilon}-factorised form with meromorphic modular
  forms}}, \href{https://doi.org/10.1007/JHEP09(2022)062}{\emph{JHEP}
  {\bfseries 09} (2022) 062},
  [\href{https://arxiv.org/abs/2207.12893}{{\ttfamily 2207.12893}}].

\bibitem{Pogel:2022ken}
S.~P\"ogel, X.~Wang and S.~Weinzierl, \emph{{The $\varepsilon$-factorised
  differential equation for the four-loop equal-mass banana graph}},
  \href{https://arxiv.org/abs/2211.04292}{{\ttfamily 2211.04292}}.

\bibitem{Dlapa:2022wdu}
C.~Dlapa, J.~M. Henn and F.~J. Wagner, \emph{{An algorithmic approach to
  finding canonical differential equations for elliptic Feynman integrals}},
  \href{https://arxiv.org/abs/2211.16357}{{\ttfamily 2211.16357}}.

\bibitem{Smirnov:2020quc}
A.~V. Smirnov and V.~A. Smirnov, \emph{{How to choose master integrals}},
  \href{https://doi.org/10.1016/j.nuclphysb.2020.115213}{\emph{Nucl. Phys. B}
  {\bfseries 960} (2020) 115213},
  [\href{https://arxiv.org/abs/2002.08042}{{\ttfamily 2002.08042}}].

\bibitem{Usovitsch:2020jrk}
J.~Usovitsch, \emph{{Factorization of denominators in integration-by-parts
  reductions}},  \href{https://arxiv.org/abs/2002.08173}{{\ttfamily
  2002.08173}}.

\bibitem{Liu:2022chg}
X.~Liu and Y.-Q. Ma, \emph{{AMFlow: a Mathematica package for Feynman integrals
  computation via Auxiliary Mass Flow}},
  \href{https://arxiv.org/abs/2201.11669}{{\ttfamily 2201.11669}}.

\bibitem{Smirnov:2015mct}
A.~V. Smirnov, \emph{{FIESTA4: Optimized Feynman integral calculations with GPU
  support}}, \href{https://doi.org/10.1016/j.cpc.2016.03.013}{\emph{Comput.
  Phys. Commun.} {\bfseries 204} (2016) 189--199},
  [\href{https://arxiv.org/abs/1511.03614}{{\ttfamily 1511.03614}}].

\bibitem{Broadhurst:1991fy}
D.~J. Broadhurst, N.~Gray and K.~Schilcher, \emph{{Gauge invariant on-shell
  Z(2) in QED, QCD and the effective field theory of a static quark}},
  \href{https://doi.org/10.1007/BF01412333}{\emph{Z. Phys. C} {\bfseries 52}
  (1991) 111--122}.

\bibitem{Melnikov:2000zc}
K.~Melnikov and T.~van Ritbergen, \emph{{The Three loop on-shell
  renormalization of QCD and QED}},
  \href{https://doi.org/10.1016/S0550-3213(00)00526-5}{\emph{Nucl. Phys. B}
  {\bfseries 591} (2000) 515--546},
  [\href{https://arxiv.org/abs/hep-ph/0005131}{{\ttfamily hep-ph/0005131}}].

\bibitem{Czakon:2007wk}
M.~Czakon, A.~Mitov and S.~Moch, \emph{{Heavy-quark production in gluon fusion
  at two loops in QCD}},
  \href{https://doi.org/10.1016/j.nuclphysb.2008.02.001}{\emph{Nucl. Phys. B}
  {\bfseries 798} (2008) 210--250},
  [\href{https://arxiv.org/abs/0707.4139}{{\ttfamily 0707.4139}}].

\bibitem{Czakon:2007ej}
M.~Czakon, A.~Mitov and S.~Moch, \emph{{Heavy-quark production in massless
  quark scattering at two loops in QCD}},
  \href{https://doi.org/10.1016/j.physletb.2007.06.020}{\emph{Phys. Lett. B}
  {\bfseries 651} (2007) 147--159},
  [\href{https://arxiv.org/abs/0705.1975}{{\ttfamily 0705.1975}}].

\bibitem{Becher:2009cu}
T.~Becher and M.~Neubert, \emph{{Infrared singularities of scattering
  amplitudes in perturbative QCD}},
  \href{https://doi.org/10.1103/PhysRevLett.102.162001}{\emph{Phys. Rev. Lett.}
  {\bfseries 102} (2009) 162001},
  [\href{https://arxiv.org/abs/0901.0722}{{\ttfamily 0901.0722}}].

\bibitem{Becher:2009qa}
T.~Becher and M.~Neubert, \emph{{On the Structure of Infrared Singularities of
  Gauge-Theory Amplitudes}},
  \href{https://doi.org/10.1088/1126-6708/2009/06/081}{\emph{JHEP} {\bfseries
  06} (2009) 081}, [\href{https://arxiv.org/abs/0903.1126}{{\ttfamily
  0903.1126}}].

\bibitem{Becher:2009kw}
T.~Becher and M.~Neubert, \emph{{Infrared singularities of QCD amplitudes with
  massive partons}},
  \href{https://doi.org/10.1103/PhysRevD.79.125004}{\emph{Phys. Rev. D}
  {\bfseries 79} (2009) 125004},
  [\href{https://arxiv.org/abs/0904.1021}{{\ttfamily 0904.1021}}].

\bibitem{Ferroglia:2009ep}
A.~Ferroglia, M.~Neubert, B.~D. Pecjak and L.~L. Yang, \emph{{Two-loop
  divergences of scattering amplitudes with massive partons}},
  \href{https://doi.org/10.1103/PhysRevLett.103.201601}{\emph{Phys. Rev. Lett.}
  {\bfseries 103} (2009) 201601},
  [\href{https://arxiv.org/abs/0907.4791}{{\ttfamily 0907.4791}}].

\bibitem{Mitov:2010xw}
A.~Mitov, G.~F. Sterman and I.~Sung, \emph{{Computation of the Soft Anomalous
  Dimension Matrix in Coordinate Space}},
  \href{https://doi.org/10.1103/PhysRevD.82.034020}{\emph{Phys. Rev. D}
  {\bfseries 82} (2010) 034020},
  [\href{https://arxiv.org/abs/1005.4646}{{\ttfamily 1005.4646}}].

\bibitem{Kidonakis:2019nqa}
N.~Kidonakis, \emph{{Soft anomalous dimensions for single-top production at
  three loops}}, \href{https://doi.org/10.1103/PhysRevD.99.074024}{\emph{Phys.
  Rev. D} {\bfseries 99} (2019) 074024},
  [\href{https://arxiv.org/abs/1901.09928}{{\ttfamily 1901.09928}}].

\bibitem{Ferroglia:2013awa}
A.~Ferroglia, S.~Marzani, B.~D. Pecjak and L.~L. Yang, \emph{{Boosted top
  production: factorization and resummation for single-particle inclusive
  distributions}}, \href{https://doi.org/10.1007/JHEP01(2014)028}{\emph{JHEP}
  {\bfseries 01} (2014) 028},
  [\href{https://arxiv.org/abs/1310.3836}{{\ttfamily 1310.3836}}].

\bibitem{Dulat:2015mca}
S.~Dulat, T.-J. Hou, J.~Gao, M.~Guzzi, J.~Huston, P.~Nadolsky et~al.,
  \emph{{New parton distribution functions from a global analysis of quantum
  chromodynamics}},
  \href{https://doi.org/10.1103/PhysRevD.93.033006}{\emph{Phys. Rev.}
  {\bfseries D93} (2016) 033006},
  [\href{https://arxiv.org/abs/1506.07443}{{\ttfamily 1506.07443}}].

\bibitem{Buckley:2014ana}
A.~Buckley, J.~Ferrando, S.~Lloyd, K.~Nordström, B.~Page, M.~Rüfenacht
  et~al., \emph{{LHAPDF6: parton density access in the LHC precision era}},
  \href{https://doi.org/10.1140/epjc/s10052-015-3318-8}{\emph{Eur. Phys. J.}
  {\bfseries C75} (2015) 132},
  [\href{https://arxiv.org/abs/1412.7420}{{\ttfamily 1412.7420}}].

\end{thebibliography}\endgroup

\end{document}